\def\mearth{{\rm\,M_\oplus}}
\def\rearth{{\rm\,R_\oplus}}
\def\msun{{\rm\,M_\odot}}

\documentclass[preprint2]{proto}
\usepackage{times}

\begin{document}

\title{\textbf{\LARGE SOLAR SYSTEM FORMATION IN THE CONTEXT OF EXTRA-SOLAR PLANETS}}

\author {\textbf{\large Sean N. Raymond}}
\affil{\small\em Laboratoire d'Astrophysique de Bordeaux, CNRS and Universit{\'e} de Bordeaux, All{\'e}e Geoffroy St. Hilaire, 33165 Pessac, France; rayray.sean@gmail.com}

\author {\textbf{\large Andre Izidoro}}
\affil{\small\em UNESP, Univ. Estadual Paulista - Grupo de Din{\^a}mica Orbital
Planetologia, Guaratinguet{\'a}, CEP 12.516-410, S{\~a}o Paulo, Brazil}

\author {\textbf{\large Alessandro Morbidelli}}
\affil{\small\em Laboratoire Lagrange, UMR 7293, Universit\'e de Nice Sophia-Antipolis,
  CNRS, Observatoire de la C\^ote d'Azur, Boulevard de l'Observatoire,
  06304 Nice Cedex 4, France}

\begin{abstract}
\begin{list}{ } {\rightmargin 1in}
\baselineskip = 11pt
\parindent=1pc
{\small 
Exoplanet surveys have confirmed one of humanity's (and all teenagers') worst fears: we are {\em weird}. If our Solar System were observed with present-day Earth technology -- to put our system and exoplanets on the same footing -- Jupiter is the only planet that would be detectable.  The statistics of exo-Jupiters indicate that the Solar System is unusual at the $\sim$1\% level among Sun-like stars (or $\sim$0.1\% among all main sequence stars). But why are we different?  

This review focuses on global models of planetary system formation. Successful formation models for both the Solar System and exoplanet systems rely on two key processes: orbital migration and dynamical instability. Systems of close-in `super-Earths' or `sub-Neptunes' cannot have formed in-situ, but instead require substantial radial inward motion of solids either as drifting mm- to cm-sized pebbles or migrating Earth-mass or larger planetary embryos. We argue that, regardless of their formation mode, the late evolution of super-Earth systems involves migration into chains of mean motion resonances anchored at the inner edge of the protoplanetary disk. The vast majority of resonant chains go unstable when the disk dissipates. The eccentricity distribution of giant exoplanets suggests that migration followed by instability is also ubiquitous in giant planet systems. We present three different models for inner Solar System formation -- the low-mass asteroid belt, Grand Tack, and Early Instability models -- each of which invokes a combination of migration and instability.  We discuss how each model may be falsified.

We argue that most Earth-sized habitable zone exoplanets are likely to form much faster than Earth, with most of their growth complete within the disk lifetime. Their water contents should span a wide range, from dry rock-iron planets to water-rich worlds with tens of percent water. Jupiter-like planets on exterior orbits may play a central role in the formation of planets with small but non-zero, Earth-like water contents. Water loss during giant impacts and heating from short-lived radioisotopes like $^{26}$Al may also play an important role in setting the final water budgets of habitable zone planets. 

Finally, we identify the key bifurcation points in planetary system formation. We present a series of events that can explain why our Solar System is so {\em weird}. Jupiter's core must have formed fast enough to quench the growth of Earth's building blocks by blocking the flux of pebbles drifting inward through the gaseous disk. The large Jupiter/Saturn mass ratio is rare among giant exoplanets but may be required to maintain Jupiter's wide orbit. The giant planets' instability must have been gentle, with no close encounters between Jupiter and Saturn, also unusual in the larger (exoplanet) context. Our Solar System system is thus the outcome of multiple unusual, but not unheard of, events. 
\\~\\~\\~}

\end{list}
\end{abstract}

\section{\textbf{INTRODUCTION}}

The discovery of extra-solar planets demonstrated that the current Solar System-inspired paradigm of planet formation was on the wrong track. Most extra-solar systems bear little resemblance to our well-ordered Solar System. While the Solar System is radially segregated, with small inner rocky worlds and more distant giant planets, few known exo-systems follow the same blueprint. Models designed with the goal of reproducing the Solar System failed spectacularly to understand why other planetary systems looked different than our own.

Yet exoplanets represent a huge sample of outcomes of planet formation, and new ideas for Solar System formation and evolution borrow liberally from models designed to explain the exoplanet population.  While we are far from a complete picture, just a handful of processes may explain the broad characteristics of most exoplanet systems and the Solar System.

We review the current thinking in how Solar System formation fits in the larger context of extra-solar planetary systems. We first (\S 1) review observational constraints on the frequency of Solar System-like systems, pointing out specific characteristics of the Solar System that don't fit within a simple formation picture. We then (\S 2) briefly summarize the stages of planet formation -- from dust to full-sized planets -- as they are currently understood, with liberal references to more detailed recent reviews of different steps.  Next (\S 3) we discuss current models for the different populations of extra-solar planets and how they match quantifiable constraints. We then turn our attention to the Solar System (\S 4). We present the empirical constraints and a rough timeline of events in Solar System formation that includes a discussion of the Nice model for the Solar System's (giant planet) dynamical instability. We discuss the classical model and its shortcomings, then present three newer competing models to match the important constraints of the inner Solar System. A challenge for all current models is to explain the mass deficit interior to Venus' orbit.  In \S 5 we extrapolate to Earth-mass planets around other stars, discussing the various formation pathways for such planets and their expected water contents. We conclude that most exo-Earths are unlikely to be truly Earth-like. Finally, in \S 6 we first synthesize these models into a large-scale picture of planetary system evolution, highlighting the key bifurcation points that may explain the observed diversity and the events that must have taken place to produce our own Solar System.  We lay out a path for future research by showing how to use theory and observations to test current models for both exoplanet and Solar System formation.  

\begin{figure*}
 \epsscale{0.95}
 \plotone{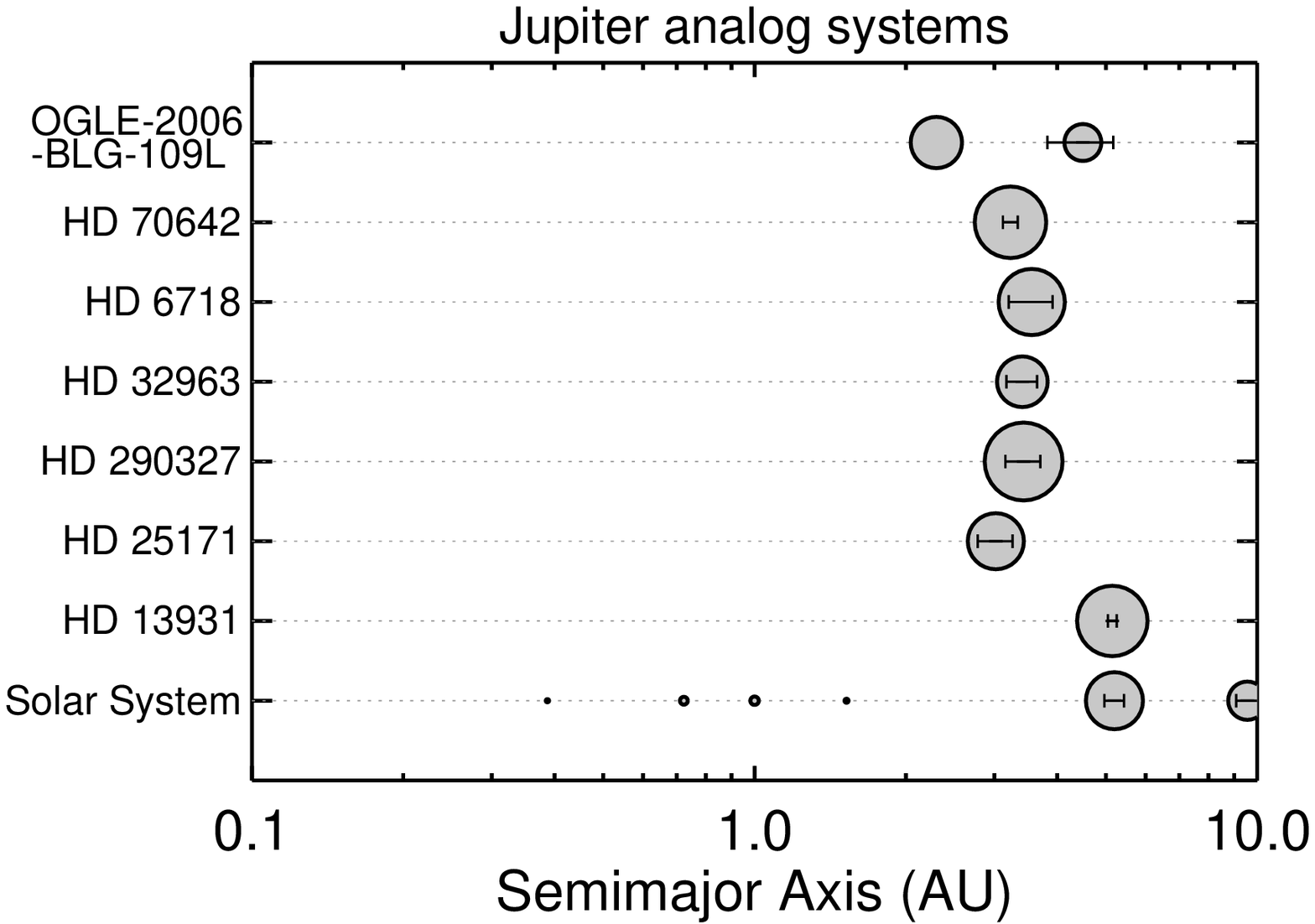}
 \caption{\small 
Seven extra-solar systems with Jupiter analogs. The top system (OGLE-2006-BLG-109L) was detected by gravitational microlensing and contains a pair of giant planets with broadly similar properties to Jupiter and Saturn orbiting a roughly half-Solar mass star~\citep{gaudi08,bennett10}.  The next six Jupiter analog systems, all detected via the radial velocity method, each have a host star within 20\% of the Sun's mass and contains a single detected planet: a gas giant with a mass between 1/3 and 3 times Jupiter's mass with a semimajor axis larger than 3 AU, and an orbital eccentricity of less than 0.1. The planet size scales with its mass$^{1/3}$ and the horizontal error bar denotes the planet's perihelion and aphelion distances. Data downloaded from exoplanets.org~\citep{wright11}. The Solar System is included for comparison. No directly-imaged planets are included in the figure, although there are two with orbits comparable in size to Saturn's: Beta Pictoris b~\citep{lagrange09} and 51 Eridani b~\citep{macintosh15}.
 }  
 \label{fig:jupanalog}
 \end{figure*}
 
\bigskip
\noindent
\textbf{1.1 How common are Solar Systems?}
\bigskip

\begin{figure}
 \epsscale{0.95}
 \plotone{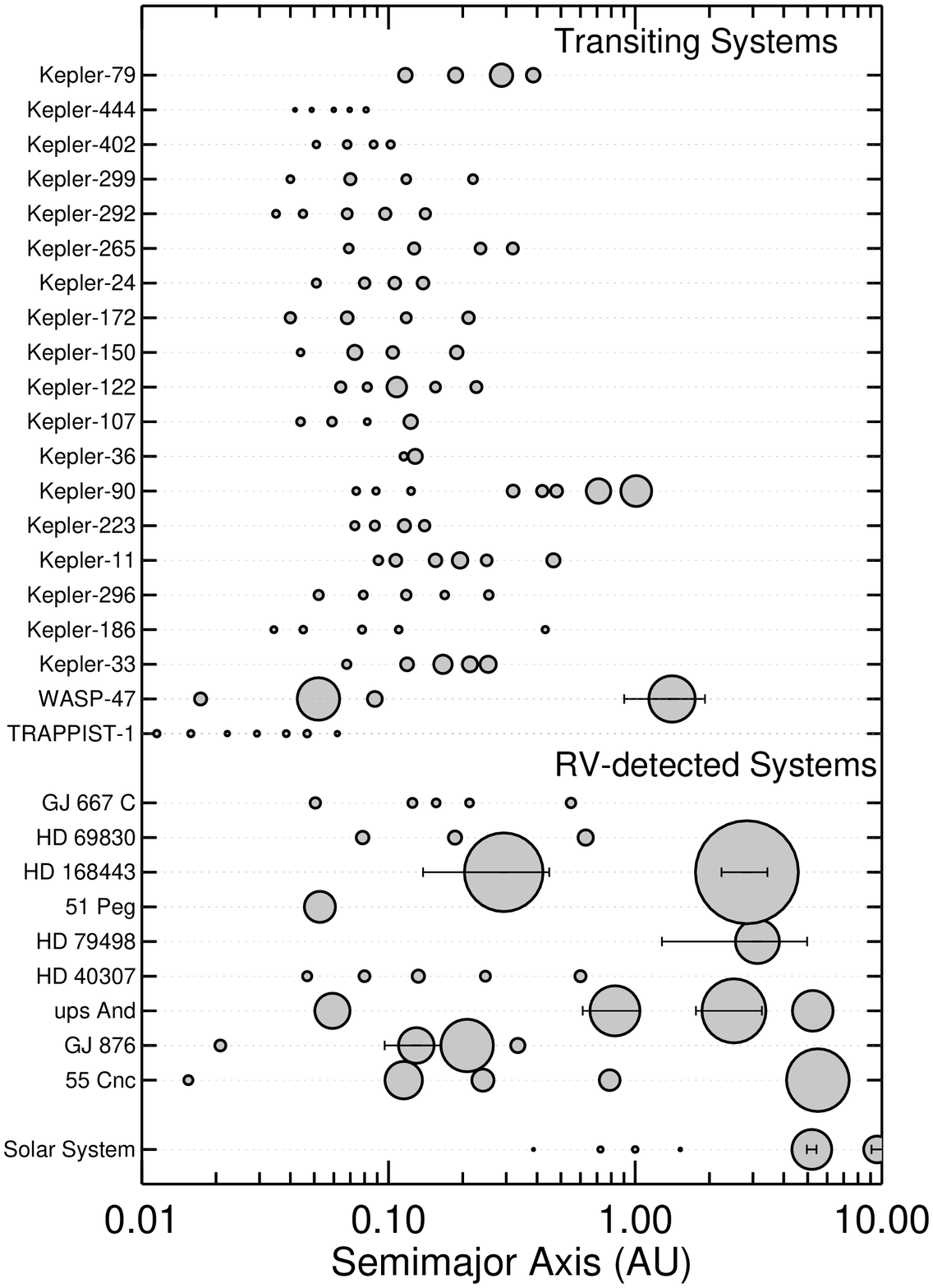}
 \caption{\small A sample of extra-solar systems chosen to illustrate their diversity (but not the true distribution of systems; for instance, most super-Earth systems contain only a single detected planet). The top systems were discovered by transit surveys and the bottom systems by radial velocity, although some planets in the transit systems also have radial velocity constraints~\citep[e.g., WASP-47;][]{sinukoff17} and some RV-detected systems host transiting planets~\citep[e.g., 55 Cnc e;][]{demory11}. The size of each planet is proportional to its true size (but not on the orbital scale); for planets with only mass (or $m \ sin \ i$) measurements, we used the $M \propto R^{2.06}$ scaling from \cite{lissauer11b}. For planets more massive than $50 \mearth$ with eccentricities larger than 0.1 (as well as for Jupiter and Saturn), the horizontal error bar represents the planet's radial excursion over its orbit (from pericenter to apocenter). Given the logarithmic x axis, the separation between adjacent planets is a measure of their period ratio regardless of their orbital radii. This plot includes a number of systems of close-in super-Earths, including two -- TRAPPIST-1~\citep{gillon17,luger17} and Kepler-223~\citep{mills16} -- in which the planets have been shown to be in long chains of orbital resonances~\citep[note that the GJ 876 system also includes a 3-planet Laplace resonance among more massive planets;][]{rivera10}. There are systems with gas giants on eccentric orbits such as Ups And~\citep{ford05}. The central stars vary dramatically in mass and luminosity for the different systems; for instance, the TRAPPIST-1 system orbits an ultracool dwarf star of just $0.08 \msun$~\citep{gillon17}. Some systems include planets that are smaller than Earth~\citep[e.g., the Kepler-444 system; ][]{campante15} and others include planets far more massive than Jupiter~\citep[e.g., HD 168443;][]{marcy01}. There are systems with roughly Earth-sized planets in their star's habitable zones, notably Kepler-186~\citep{quintana14}, TRAPPIST-1~\citep{gillon17}, and GJ 667 C~\citep{anglada13}. Some of these planets are in multiple star systems~\citep[e.g., 55 Cnc;][]{fischer08}.}  
 \label{fig:interesting}
 \end{figure}

To date, radial velocity (RV) and transit surveys have discovered thousands of extra-solar planets. Figure~\ref{fig:interesting} shows a sample of the diversity of detected exo-systems. These surveys have determined occurrence rates of planets as a function of planet size/mass and orbital period around different types of stars~\citep{howard10,howard12,mayor11,fressin13,dong13,petigura13,fulton17}.  Meanwhile, gravitational microlensing and direct-imaging surveys have placed constraints on the properties of outer planetary systems~\citep{cassan12,biller13,mroz17,bowler18}.\footnote{Microlensing may actually be the most sensitive method for detecting analogs to our Solar System's giant planets. Indeed, microlensing observations have found a Jupiter-Saturn analog system~\citep{gaudi08,bennett10} as well as rough analogs to the ice giants~\citep{poleski14,sumi16}. However, given that microlensing requires a precise alignment between a background source and the star whose planets can be found~\citep[e.g.][]{gould92}, it cannot be used to search for planets around a given star.  Rather, its power is statistical in nature~\citep[see][]{gould10,clanton14,clanton16,suzuki16}. Nonetheless, upcoming microlensing surveys -- especially space-based surveys such as WFIRST -- are expected to find hundreds to thousands of planets in the Jupiter-Saturn regions of their stars~\citep{penny18}.}

To put the Solar System on the same footing as the current sample of extra-solar planets we must determine what our system would look like when observed with present-day Earth technology. The outcome is somewhat bleak. The terrestrial planets are all too small and too low-mass to be reliably detectable.  Although sub-Earth-sized planets were discovered by {\it Kepler}~\citep[e.g.][]{barclay13}, and $\sim$Earth-mass planets have been found by radial velocity monitoring~\citep[e.g.][]{anglada16}, they were all on close-in orbits.  Strong observational biases make it extremely challenging to detect true analogs to our terrestrial planets~\citep[e.g.,][]{charbonneau07,fischer14,winn18}.  However, a decade-long radial velocity survey would detect Jupiter orbiting the Sun.  Indeed, several Jupiter analogs have been discovered~\citep[e.g.][]{wright08}.  Saturn, Uranus and Neptune are too distant to be within the reach of radial velocity surveys.  Figure~\ref{fig:jupanalog} shows seven known Jupiter analog systems for scale.

The exo-Solar System is therefore just the Sun-Jupiter system. Observations of the Solar System as an exoplanet system would provide a decent measurement of Jupiter's mass (really, its $m \ sin\ i$, where $i$ is the angle between our line of sight and its orbital plane) and semimajor axis, with modest constraints on its orbital eccentricity. 

Based on current data, the Sun-Jupiter system is rare at the one-in-a-thousand level. The Sun -- a member of the G dwarf spectral class -- is much more massive than most stars; among nearby stars only $\sim$5\% have similar masses~\citep[e.g.][]{chabrier03}. Roughly 10\% of Sun-like stars have gas giant planets~\citep[defined as having masses $M \gtrsim 50 \mearth$; ][]{butler06,udry07b,cumming08,mayor11,clanton14}.  However, most have orbits that are either significantly closer-in or more eccentric. Using a relatively broad definition for Jupiter-like planets -- as planets with orbital radii larger than 2 AU and orbital eccentricities below 0.1 -- only 10\% of giant exoplanets are Jupiter-like. This puts Jupiter as a 1\% case among Sun-like stars, or $\sim 0.1\%$ overall.  

The Solar System's peculiarity can also be considered in terms of planets that are present in other systems but absent in ours.  At least 30-50\% of main sequence stars have planets smaller than $4 \rearth$ (or less massive than $\sim 10-20 \mearth$) on orbits closer-in than Mercury's~\citep{mayor11,howard12,fressin13,petigura13,dong13,hsu18,zhu18}. Recent modeling suggests that less than 8\% of planetary systems have their innermost planet on an orbit as wide as Mercury's, and less than 3\% have an innermost planet on an orbit as wide as Venus'~\citep{mulders18}.  This reinforces the Solar System's standing as a outsider.

The Solar System's relative scarcity among exoplanet systems falls at an interesting level.  We are not so rare that no Solar System analogs have been found.  Nor are we so common to be just a ``face in the crowd'' among exoplanets.  Based on a single detectable planet, the Solar System stands apart from the crowd but not alone.  The question is, why?

\bigskip
\noindent
\textbf{1.2 Peculiarities of the Solar System}
\bigskip

The orbital architecture of the Solar System presents a number of oddities. But like a polka lover's musical preferences, these oddities only become apparent when viewed within a larger context. The ``classical model'' (discussed in \S 4.3) offers a convenient reference frame for the origin of the terrestrial planets. The classical model assumes that the planets formed mainly in-situ, meaning from building blocks that originated at roughly their current orbital distances. It also assumes that terrestrial- and giant planet formation can be considered separately. 

The classical model invokes bottom-up planetary accretion. Starting from a distribution of solids, the planets that form retain a memory of their initial conditions~\citep[e.g.][]{raymond05}. This motivated the `minimum-mass solar nebula' model~\citep{weidenschilling77,hayashi81}, which uses the planets' present-day orbits to reconstruct a disk from which they may have formed (neglecting any significant radial motion like orbital migration). This style of growth leads to systems in which planets on adjacent orbits have similar sizes~\citep[e.g.][]{kokubo02}. The observed super-Earths do appear to have similar sizes within a given system~\citep{millholland17,weiss18}, although it is hard to imagine a scenario for their formation that does not invoke large-scale radial drift of solids (see \S 3.2).  

When we compare the classical model blueprint with the actual Solar System, several discrepancies emerge:
\begin{itemize}
\item {\bf Why is Mars so much smaller than Earth?}  Simulations of terrestrial planet formation from a smooth disk tend to produce Earth and Mars analogs with similar masses~\citep[e.g.][]{wetherill78,chambers01,raymond06b}, in contrast with the actual 9:1 mass ratio between the two planets. This is called the `small Mars' problem, which was first pointed out by \cite{wetherill91} and has motivated a number of models of terrestrial planet formation (see \S 4).
\item {\bf Why is Mercury so much smaller than Venus?} Although it receives far less attention, the large Venus/Mercury mass ratio is an even bigger problem than the `small Mars' problem. The Venus/Mercury mass ratio is 14:1 but simulations again tend to produce planets with similar masses and with more compact orbital configurations than the real one. It is worth noting that, in the context of extra-solar planetary systems (in which super-Earths are extremely common), the mass deficit interior in the very inner Solar System is in itself quite puzzling.
\item {\bf Why are the asteroid and Kuiper belts so low-mass yet dynamically excited?} The asteroid and Kuiper belts contain very little mass: just $\sim 5 \times 10^{-4} \mearth$ and $\sim 0.01-0.1 \mearth$, respectively~\citep{demeo13,gladman01}. However, both belts are dynamically excited, with much higher eccentricities and inclinations than the planets. Yet the mass required to self-excite those belts exceeds the present-day mass by many orders of magnitude~\citep[e.g.,][]{obrien07}. This apparent contradiction is a key constraint for Solar System formation models.

\item {\bf Why is Jupiter's orbit so wide, and why aren't all giant exoplanets in orbital resonance?}  Our giant planets present apparent contradictions when viewed through the lens of orbital migration.  Migration is an inevitable consequence of planet formation. Given that planets form in massive gaseous disks, gravitational planet-disk interactions {\em must} take place. Migration is generally directed inward and the co-migration of multiple planets generically leads to capture in mean motion resonances~\citep[e.g.,][]{kley12,baruteau14}.  

\end{itemize}


\section{\textbf{STAGES OF PLANET FORMATION}}

Global models of planet formation can be thought of as big puzzles. The puzzle pieces are the processes involved in planet formation, shaped by our current level of understanding.  We now briefly review the stages and processes of planetary formation as envisioned by the current paradigm.  We remain brief and refer the reader to recent reviews for more details.

{\bf Protoplanetary disks}.  While high-resolution observations of disks around young stars show exquisite detail~\citep[e.g.,][]{alma15,andrews16}, the structure and evolution of the dominant, gaseous component of planet-forming disks remains uncertain~\citep[see discussion in][]{morbyraymond16}. Observations of disk spectra suggest that gas accretes from disks onto their stars~\citep{meyer97,hartmann98,muzerolle03}, and the occurrence rates of disks around stars in clusters of different ages suggest that disks dissipate on a few million year timescale~\citep{haisch01,briceno01,hillenbrand08b,mamajek09}. Disk models thus depend on mechanisms to transport angular momentum in order to generate large-scale radial gas motion~\citep{balbus98,turner14,fromang17}. Historically, models have assumed that disks are sufficiently ionized for the magneto-rotational instability to generate viscosity across the disk~\citep{lyndenbell74} often using the so-called alpha prescription~\citep{shakura73}.  However, recent models including the Hall effect and ambipolar diffusion terms have found a fundamentally different structure and evolution than alpha-disks~\citep{lesur14,bai16,suzuki16}, and this structure has implications for multiple stages of planet formation and migration~\citep{morbyraymond16}. The final dissipation of the disk is thought to be driven by photo-evaporation from the central star~\citep[and in some cases by external UV field;][]{hollenbach94,adams04}. \citep[For reviews of disk dynamics, structure and dispersal, see][]{armitage11,turner14,alexander14,ercolano17}).  

{\bf From dust to planetesimals.} Based on observed infrared excesses in, sub micron-sized dust particles are observed to be very abundant in young protoplanetary disks~\citep[e.g.][]{briceno01,haisch01}. Dust particles growing by coagulation in a gaseous disk encounter a number of barriers to growth such as fragmentation and bouncing~\citep{brauer08,guttler10}. Once they reach roughly mm-size (or somewhat larger) particles very rapidly drift inward, leading to what was historically called the ``meter-sized barrier'' to growth~\citep{weidenschilling77b,birnstiel12}. New models suggest that, if they are initially sufficiently concentrated relative to the gas, the streaming instability can produce clumps of drifting particles that are bound together by self-gravity and directly form 100 km-scale planetesimals~\citep{youdin05,johansen09,johansen15,simon16,yang17}. Planetesimals are the smallest macroscopic bodies that do not undergo rapid aerodynamic drift, and are often considered the building blocks of planets. Exactly where and when planetesimals form depends itself on the dynamics and structure of the disk~\citep{drazkowska16,carrera17}, and some recent studies suggest that planetesimal growth may be favored near the inner edge of the disk~\citep{drazkowska16} and just past the snow line~\citep{armitage16,drazkowska17,schoonenberg17}. 
\citep[For reviews of dust growth/drift and planetesimal formation, see:][]{blum08,chiang10,johansen14,birnstiel16}.  

{\bf Pebble- and planetesimal accretion.} Planetesimals can grow by accreting other planetesimals~\citep{greenberg78,wetherill93,kokubo00} or pebbles that continually drift inward through the disk~\citep{ormel10,johansen10,lambrechts14}. Pebbles are defined as particles for which the gas drag timescale is similar to the orbital timescale, and are typically mm- to cm-sized in the terrestrial- and giant planet-forming regions of disks~\citep[see][]{johansen17}. Pebbles are thought to continually grow from dust and drift inward through the disk, such that growing planetesimals see a radial flux of pebbles across their orbits~\citep{lambrechts14b,chambers16,ida16}. At low relative speeds, a large planetesimal efficiently accretes nearby small particles (either pebbles or small planetesimals) because the large planetesimal's gravity acts to increase its effective collisional cross section~\citep[a process known as gravitational focusing; ][]{safronov69,rafikov04,chambers06}. This triggers a phase of runaway growth~\citep{greenberg78,wetherill93,kokubo98}. At later stages, growth by the accretion of other planetesimals is self-limited because the growing planetesimal excites the random velocities of nearby planetesimals, decreasing the efficiency of gravitational focusing~\citep{kokubo00,leinhardt05}. However, gas drag acts much more strongly on pebbles and maintains their low velocities relative to larger bodies. The efficiency of pebble accretion increases with the growing planetesimal's mass~\citep{lambrechts12,morby12}, and pebble accretion outpaces planetesimal accretion for bodies more massive than roughly a lunar mass~\citep[$0.012 \mearth$; although the exact value depends on the parameters of the disk;][]{johansen17}.  Above roughly a lunar mass these objects are generally referred to as ``planetary embryos''. When an embryo reaches a critical mass called the {\it pebble isolation mass} it generates a pressure bump in the disk exterior to its orbit, which acts to block the inward flux of pebbles~\citep{morby12,lambrechts14b,bitsch18}. This acts to quench not only the embryo's growth but also the growth by pebble accretion of all objects interior to the embryo's orbit. Later growth must therefore rely on the accretion of planetesimals, other embryos or gas. We note that there is some debate about whether pebble accretion remains efficient for planets with significant gaseous envelopes below the pebble isolation mass~\citep[see][]{alibert17,brouwers18}. \citep[For reviews of pebble- and planetesimal accretion, see:][respectively]{johansen17,kokubo02}

{\bf Gas accretion and giant planet growth.} Once planetary embryos become sufficiently massive they accrete gas directly from the disk~\citep{pollack96,ida04,alibert05}. Gas accretion operates to some degree for Mars-mass planetary embryos, and there is evidence from noble gases in Earth's atmosphere that a portion of the atmospheres of Earth's constituent embryos was retained during Earth's prolonged accretion~\citep{dauphas03}.  Gas accretion onto a growing planet depends on the gaseous envelope's opacity~\citep{ikoma00,hubickyj05,machida10} and temperature, which is determined in part by the accretion rate of energy-depositing solid bodies~\citep{rice03,broeg12}. The dynamics of how gas is accreted onto a growing planet's surface is affected by small-scale gas flows in the vicinity of the planet's orbit~\citep{fung15,lambrechts17} as well as the structure of the circum-planetary disk~\citep[if there is one; e.g., ][]{ayliffe09,szulagyi16}. When the mass in a planet's envelope is comparable to its solid core mass it undergoes a phase of runaway gas accretion during which the planet's expanding Hill sphere -- the zone in which the planet's gravity dominates the star's -- puts it in contact with ever more gas, allowing it to grow to a true gas giant on the Kelvin-Helmholtz timescale~\citep{mizuno80,ida04,thommes08a}. This culminates with the carving of an annular gap in the disk, which slows the accretion rate~\citep{bryden99,crida06,lubow06}. 
\citep[For reviews of giant planet growth, see:][]{lissauer07,helled14}

{\bf Orbital migration.} Migration is an inevitable consequence of gravitational interactions between a growing planet and its natal gas disk. Planets launch density waves in the disk, whose flow is determined by the disk's dynamics. These density perturbations impart torques on the planets' orbits~\citep{goldreich80,ward86}. These torques damp planets' eccentricities and inclinations~\citep{papaloizou00,tanaka04} and also drive radial migration. For planets low enough in mass not to carve a gap in the disk (i.e., for planetary embryos and giant planet cores) the mode of migration is sometimes called {\it type I}. Multiple torques are at play. The {\it differential Lindblad torque} is almost universally negative, driving planets inward at a rate that is proportional to the planet mass~\citep{ward97,tanaka02}. In contrast, the {\it corotation torque} can in some situations be positive and overwhelm the differential Lindblad torque, leading to outward migration~\citep{paardekooper06,kley08,masset10,paardekooper11,benitez15}. The disk structure plays a central role in determining the regions in which outward migration can take place~\citep{bitsch13,bitsch15}. Planets that carve a gap enter the {\it type II} migration regime, which is generally slower than type I. In this regime a planet's migration is determined in large part by the radial gas flow within the disk and thus its viscosity~\citep{ward97,durmann15}. Type II migration is directed inward in all but a few situations~\citep{masset01,veras04,crida09,pierens14}.
\citep[For reviews of migration, see: ][]{kley12,baruteau14}

{\bf Late-stage accretion}. The gaseous component of planet-forming disks is observed to dissipate after a few million years~\citep{haisch01,hillenbrand08b}. When the gas disk is gone gas accretion, pebble accretion, gas-driven migration and dust drift cease, and the final phase of growth begins. Late-stage accretion consists of a protracted phase of collisional sweep-up of remaining planetary embryos and planetesimals~\citep{wetherill78,wetherill96,chambers01,raymond04}. Whereas some late accretion should occur for all planets, this process is thought to have played a predominant role in the growth of Earth and Venus. For these planets, this phase was characterized by giant impacts between Mars-mass or larger planetary embryos~\citep[e.g.,][]{agnor99,stewart12,quintana16}. The very last giant impact on Earth (and perhaps in the whole inner Solar System) is thought to have been the Moon-forming impact~\citep{benz86,canup01,cuk12}.  Dynamical friction from remnant planetesimals -- which leads to an energy equipartition between planetesimals and embryos -- keeps the orbits of embryos relatively circular during the early parts of late-stage accretion~\citep{obrien06,raymond06b}. While dynamical friction dwindles in importance as the planetesimal population is eroded by impacts and dynamical clearing, it is nonetheless important in setting the planets' final orbits. Growth generally proceeds inside to out.  The timescale for Solar System-like systems to complete the giant impact phase is $\sim$100 Myr~\citep{chambers01,raymond09c,jacobson14}.  Analogous terrestrial planet systems around different types of stars would have different durations of this phase~\citep{raymond07b,lissauer07b}. However, some systems (e.g., those in resonant chains) may never undergo a late-stage phase of giant impacts. \citep[For reviews of late-stage accretion see:][]{morby12b,raymond14,izidoro18}

\section{\textbf{EXOPLANET FORMATION MODELS}}

We now review formation models for extra-solar planets.  We focus on two specific categories of exoplanets: super-Earth systems and giant exoplanets.  For each category we summarize the observational constraints, then present the relevant models.  It is interesting to note that, despite the very different regimes involved, leading models for each category rely heavily on two processes: migration and dynamical instability.

\bigskip
\noindent
\textbf{3.1 Systems of close-in low-mass planets (`super-Earths')}
\bigskip

The abundance of close-in low-mass/small planets is one of the biggest surprises to date in exoplanet science. Both radial velocity and transit surveys find an occurrence rate of 30-50\% for systems of planets with $R < 4 \rearth$ or $M < 10-20 \mearth$ and $P < 50-100$~days~\citep[see][for a compilation of measured rates and a comparison between them]{winn15}.  These planets are commonly referred to as `super-Earths', even though a significant fraction appear to be gas-rich and thus closer to `sub-Neptunes'~\citep{rogers15,wolfgang16,chen17}.

Super-Earth formation models are constrained by several lines of observations: 
\begin{itemize}
\item Their high occurrence rate~\citep{howard10,howard12,mayor11,fressin13,dong13,petigura13}, and the stellar mass-dependence of their properties. Low-mass stars appear to have a similar overall abundance of close-in small planets but, compared with Sun-like stars, they have more super-Earths (with $R<2\rearth$), fewer sub-Neptunes (with $R>2\rearth$) and a higher average total mass in planets~\citep{dressing15,mulders15,mulders15b}.
\item The distribution of orbital period ratios of adjacent planets~\citep{lissauer11b,fabrycky14,steffen15}. Only a small fraction (perhaps 5-10\%) of pairs or neighboring planets appear to be in resonance.
\item The multiplicity distribution, or how many planets are detected around each star~\citep{lissauer11b,batalha13,rowe14,fabrycky14}. There is a large peak in the transit detection of singleton super-Earth systems compared with multiple planet systems~\citep[the so-called {\it Kepler dichotomy}; see, e.g., ][]{johansen12,fang12,ballard16}. It remains debated whether this `dichotomy' is a signature of planet formation or simply an observational bias~\citep[e.g.][]{izidoro17,izidoro18b,zhu18}.
\item The distribution of planet sizes/mass and the size/mass ratios of adjacent planets.  Recent analysis of the Kepler super-Earths has found that adjacent planets tend to be similar-sized~\citep{millholland17,weiss18}, consistent with the classical model but at odds with our own terrestrial planets.
\item The distribution of physical densities of planets.  Densities can be measured for transiting planets with good mass constraints from radial velocity monitoring~\citep[e.g.][]{fischer14,marcy14} or transit-timing variation analysis~\citep[see][for a review]{agol17}. Several analyses based on models of planetary interiors have concluded that small super-Earths are predominantly rocky whereas large super-Earths to have thick gaseous envelopes, with a transition between the two regimes somewhere in the range of $1.2-2 \rearth$~\citep{rogers15,wolfgang16,chen17}. 
\end{itemize}

All of these constraints are naturally subject to observational bias, making a comparison with models challenging.  For example, the period ratio distribution could be skewed if the middle planet in a 3-planet system is not detected, leading to a detection of periods $P_3$ and $P_1$ but not $P_2$ and hence an inflated period ratio $P_3/P_1$.  Alternately, the inferred distribution could be skewed by preferentially missing the transit of outer planets in pairs with large period ratios, i.e. by detecting $P_2/P_1$ but not $P_3/P_2$ if $P_3/P_2$ is much larger than $P_2/P_1$~\citep[see discussion in][]{izidoro18b}.

A number of formation models for super-Earths existed prior to their discovery. \cite{raymond08a} compiled six different formation mechanisms for super-Earths (although in that paper they were referred to as ``hot Earths''). Many of those mechanisms relied on the influence of gas giants in the system~\citep[e.g.][]{fogg05,zhou05,raymond06b,mandell07} and are thus unable to match the general population of super-Earths (although those mechanisms could apply in select cases).  

The simplest model of super-Earth formation -- in-situ accretion -- was proposed by \cite{raymond08a} who subsequently discarded it as unrealistic~\citep[see also][]{ogihara15}.  Given its recent revival~\citep{chiang13,hansen12,hansen13} and surprising popularity in the exoplanet community we think it worth explaining exactly why it cannot be considered a viable model. The simplest argument against in-situ accretion is simply that it is not self-consistent. If super-Earths accreted in-situ close to their stars, then their natal planet-forming disks must have been quite dense~\citep{raymond08a,chiang13,schlichting14,schlaufman14}. The timescales for accretion of these planets are very short because their disks are massive and the orbital time is short~\citep{safronov69}.  Simulations demonstrate that planets similar to the observed super-Earths indeed accrete on thousand- to hundred-thousand-year timescales~\citep{raymond08a,hansen12,bolmont14}, long before the dispersal of gaseous disks. Thus, super-Earths must have been massive enough to gravitationally interact with the gaseous disk, which itself must have been extremely dense to accommodate the planets' in-situ growth. The planets must therefore have migrated~\citep{ogihara15}. In fact, the disks required for in-situ accretion are so dense that even aerodynamic drag alone would have caused their orbits to shrink~\citep{inamdar15}. 

The in-situ model is thus caught in a logical impossibility. If the planets formed in-situ then they must have migrated. But if they migrated, they did not form in-situ. In other words, super-Earths simply cannot have formed in-situ. Solids must have drifted relative to the gas, either at small scales (pebble drift) or large scales (migration). Meanwhile, there is abundant circumstantial evidence for inward planet migration. One extreme case is the existence of planets interior to the silicate sublimation radius~\citep{swift13}. Another piece of evidence is the existence of systems of super-Earths in resonant chains in systems such as Kepler-223~\citep{mills16} and TRAPPIST-1~\citep{gillon17,luger17}, given that it is extremely improbable for planets to end up in low-eccentricity resonant configurations without invoking migration.

Two models remain viable candidates to explain the origin of most super-Earths: the {\it drift} and {\it migration} models.  

The drift model~\citep{boley13,chatterjee14,chatterjee15,hu16,hu17} proposes that inward-drifting pebbles are trapped in the inner parts of the disk, perhaps at a pressure bump associated with a region toward the inner edge of the disk where its properties (e.g., its viscosity) change abruptly.  Pebbles accumulate at the pressure bump until a threshold is reached for them to form planetesimals~\citep[e.g.][]{yang17} and accrete into full-sized planets. \cite{chatterjee14} proposed that the pressure bump itself would respond to the first planet's presence and retreat to an external orbit, providing a new nexus of super-Earth formation.  While the model's predictions appear broadly consistent with observations, it has not been developed to the point of matching the observables laid out above. Below we argue that the late evolution of super-Earths in this model must also include migration. 

A number of studies have placed themselves at the interface between the in-situ accretion and drift models~\citep[e.g.][]{hansen12,hansen13,dawson15,dawson16,lee16,lee17,moriarty16}. These studies assumed that the bulk of super-Earths' accretion happens close-in but within disks that have far less gas than would be present assuming that the local density of gas reflects the local density of solids (e.g., with a roughly 100 to 1 ratio, assuming Solar metallicity). These studies thus inherently suppose that a previous process of solid enrichment took place within the inner disk, presumably by a mechanism such as dust/pebble drift. They further assume that planetesimals were uniformly distributed across the inner disk, usually as a simple, power-law profile. Such initial conditions -- young disks with a broad close-in planetesimal distribution but little gas -- are hard to reconcile with current thinking.  If planetesimals formed quickly then the gas density should still be high and migration should be very fast.  If planetesimals formed more slowly, presumably from pebbles that drifted inward to supplement the inner disk's solid reservoir, then they are unlikely to have a smooth radial distribution~\citep[e.g.][]{chatterjee14,drazkowska16}.  Thus, while these studies have provided interesting insights into various aspects of the accretion process, it seems unlikely that their starting conditions reflect reality.  The mechanism by which inner disks are enriched in solids seems to be central to understanding the origins of super-Earths.  

The migration model~\citep{terquem07,ogihara09,mcneil10,ida10,cossou14,izidoro17,izidoro18b} proposes that large planetary embryos form throughout the disk and migrate inward. Inward migration is counteracted by the positive surface density gradient at the disk's inner edge~\citep{masset06}.\footnote{The disk inner edge thus provides a built-in stopping mechanism for inward migration that plays a central role in the migration model. One may then wonder whether the in-situ formation model could also have a mechanism for avoiding migration. Indeed, in-situ growth within a region of slow or stopped migration -- such as near the disk's inner edge -- would naturally reduce the importance of migration. However, there are two important caveats.  First, if planets grew in-situ in slow-migration regions, this would not remove the importance of planet-disk interactions, which also affect the growing planets' eccentricities and inclinations~\citep[e.g.][]{papaloizou00,tanaka04}. Second, regions of reduced migration are thought to be narrow~\citep[e.g.][]{hasegawa11,bitsch15,baillie15} such that even if some super-Earths did indeed grow more or less in-situ, the bulk of growing super-Earths would still have experienced migration.} Embryos thus migrate inward and pile up into long resonant chains anchored at the disk's inner edge. Collisions are common during this phase, leading to a breaking of resonance followed by continued migration and rapid re-formation of the resonant chain in a new configuration. After the gas disk dissipates many resonant chains become unstable, leading to a phase of giant impacts between growing planets that is not unlike the final phase of in-situ accretion. If 90-95\% of resonant chains become unstable, the resulting systems provide a quantitative match to the observed period ratio and multiplicity distributions~\citep{izidoro17,izidoro18b}. In this model, the Kepler dichotomy is an observational artifact: the broad distribution of mutual inclinations in multiple super-Earth systems naturally produces a peak in systems with a single transiting planet. The resonant chains that remain stable are associated with observed multi-resonant systems such as TRAPPIST-1~\citep{gillon17,luger17} and Kepler-223~\citep{mills16}.  

\cite{ormel17} proposed a hybrid scenario in which planetesimals form first at the snow line, undergo pebble accretion and then migrate inward. This idea is consistent with the migration model and also connects with dust growth and drift models, which find that planetesimals tend to form fastest just past the snowline~\citep{armitage16,drazkowska17,schoonenberg17}. 

The late stages in super-Earth evolution should be the same for both the drift and migration models. Of course, the two models invoke different formation modes and feeding zones for the planets.  However, once there is a population of planets massive enough to migrate, the subsequent evolution is independent of how the planets formed. Whatever the processes responsible for creating a population of such planets, they migrate. And the outcome of this migration is a well-studied problem. As long as there is a disk inner edge, a system of migrating planets invariably organizes itself into a chain of mean motion resonances~\citep[e.g.][]{snellgrove01,lee02,papaloizou06,cresswell07}. Magnetohydrodynamic simulations of disk accretion onto young stars show that for most plausible parameters disks should indeed have inner edges~\citep{romanova03,bouvier07}. Although the strength of the (positive) corotation torque depends on the local disk properties~\citep{masset10,paardekooper11}, migration is likely to be directed inward during the late phases of disk evolution~\citep{bitsch14,bitsch15}. Another factor is that a system of many super-Earths in resonance acts to excite their mutual eccentricities, decreasing the strength of the corotation torque~\citep{bitsch10,fendyke14} and potentially leading to inward migration of the cohort~\citep{cossou13}. 

We therefore argue that super-Earths must migrate regardless of how they formed. Late in the disk lifetime all formation scenarios converge on the evolution envisioned in the ``breaking the chains'' model of \cite{izidoro17,izidoro18b}. Migration should produce resonant chains with the innermost planets anchored at the inner edge of the disk. When the gas disk dissipates along with its stabilizing influence, the vast majority of resonant chains become unstable. While many details remain to be resolved, this evolution matches the key observed super-Earth constraints. 

\cite{raymond08a} proposed that super-Earth formation models could be differentiated with two observables: the planets' compositions (via their densities) and the systems' orbital architectures.  We have just argued that the late phases of the two viable models should converge to the same dynamical pathway, i.e., the ``breaking the chains'' model. We therefore do not expect the orbital architectures of super-Earth systems to provide a means of differentiation between models.

The compositions of super-Earths should in principle be different for the drift and migration models. In the drift model, all planet-building takes place close-in. Super-Earths should therefore be purely rocky, because temperatures so close-in are so hot that the local building blocks should not contain any volatiles such as water.  In contrast, a simple view of the migration model would predict ice-rich planets. Indeed, some models of dust growth and drift first produce planetesimals at or beyond the snow line, the radial distance beyond which water vapor can condense as ice~\citep{armitage16,drazkowska17}. Immediately past the snow line, embryos are thought to grow faster and larger than closer-in because of pebbles may be somewhat larger and therefore easier to accrete, and may also be efficiently concentrated~\citep{ros13,morby15}. This would suggest that ice-rich embryos from past the snow line are likely to be the first to migrate and thus, super-Earths should themselves be ice-rich in the migration model.

There are three problems with the idea of the migration model producing exclusively ice-rich super-Earths~\citep[see][]{raymond18b}.  First, in some cases rocky embryos may grow large enough to migrate~\citep[see][]{lambrechts18,izidoro18b}. Indeed, in some models planetesimals form first in the terrestrial planet-forming region~\citep{drazkowska16,surville16}, which would give rocky embryos a head-start in their growth. Second, migrating embryos must pass through the building blocks of terrestrial planets on their way to becoming super-Earths. Their migration can act to pile up rocky material in inner resonances with the migrating embryos~\citep{izidoro14} and catalyze the rapid formation of rocky super-Earths, which preferentially end up interior to the migrating ice-rich embryos. Indeed, simulations of the migration model show that the innermost super-Earths are often built entirely from inner planetary system material and should be purely rocky~\citep{raymond18b}. Third, it is not clear that embryos that migrate inward from beyond the snow line must be ice-rich. It is the fastest-forming embryos that are most likely to migrate, for simple timescale reasons. Yet rapid growth implies massive volatile loss. Thermal evolution models find that any planetesimals that form within 1 Myr are completely dehydrated by strong $^{26}$Al heating~\citep{grimm93,monteux18}. This short-lived radionuclide (half-life of $\sim$700,000 years) is thought to have been injected into the Sun's planet-forming disk from a nearby massive star~\citep[e.g.][]{hester04,gounelle08,gaidos09,oullette10} and to have played a central role in the thermal evolution of the fastest-forming planetesimals. In addition, giant collisions between ice-rich bodies preferentially strip outer icy mantles and leave behind rocky/iron cores~\citep{marcus10}. Likewise, later giant impacts can lead to substantial water loss for ocean planets~\citep{genda05}.

It is the very closest-in planets that are easiest to characterize. Within the so-called ``photo-evaporation valley'', the atmospheres of any super-Earth-sized planets are thought to be rapidly evaporated by UV irradiation from the central star~\citep{lammer03,baraffe04,hubbard07a}. Planets in this region should necessarily have lost their gaseous envelopes and thus be ``naked''~\citep{lopez13,owen13,sanchisojeda14}, while planets just past the valley may have larger radii due to atmospheric heating~\citep{carrera18}. From their observed sizes and masses, these very close-in planets appear to be rocky in nature, not icy~\citep{owen17,lopez17,jin18}. Some studies have suggested that the observed dip in planet occurrence between $\sim 1.5-2 \rearth$~\citep{fulton17,teske18,fulton18} may provide evidence that super-Earths are rocky~\citep{jin18,vaneylen18}. \cite{gupta18} suggest that in this context, `rocky' planets should have less than 20\% water by mass (which is comparable to Europa's water content by mass). Given that both the migration and drift models are consistent with rocky super-Earths~\citep{raymond18b}, observations cannot yet differentiate between the drift and migration models.  Interpreting the mean densities of planets beyond the distance for atmospheric photo-evaporation is challenging because of the degeneracies that arise once gas is included as a third potential constituent~\citep[along with rock/iron and water; see][]{selsis07b,adams08}. The main difference between the migration and drift models is that the migration model predicts that some super-Earths should be ice-rich, in particular those planets that formed relatively late or in disks with little $^{26}$Al.

Embryos embedded in the disk should also accrete gas~\citep[e.g.][]{ikoma00,rogers11}. Indeed, many super-Earths are observed to have very low densities~\citep[e.g.][]{marcy14} and this has been interpreted as planets larger than $\sim 1.5 \rearth$ having gaseous envelopes that are typically 0.1-10\% of their total mass~\citep{lopez14,weiss14,rogers15,wolfgang16,chen17}. 

\begin{figure*}
 \epsscale{1.95}
 \plotone{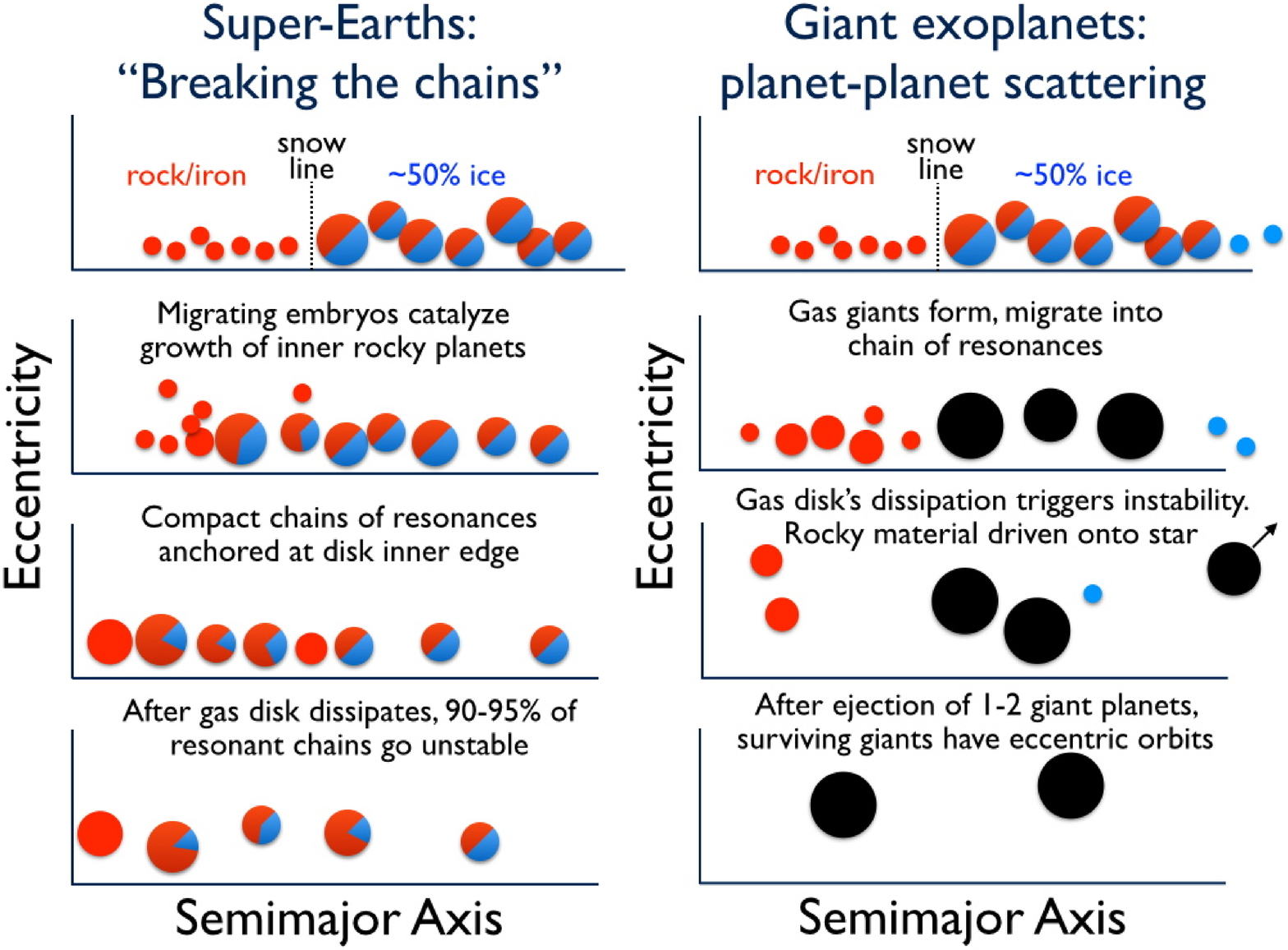}
 \caption{\small How orbital migration and dynamical instabilities can explain the properties of exoplanet populations. {\bf Left:} Evolution of the ``breaking the chains'' migration model for the origin of super-Earths~\citep{izidoro17,izidoro18b}.  Embryos within the snow line are entirely rocky and much smaller than those that form past the snow line, which also incorporate ice. Presumably ice-rich embryos migrate inward through the rocky material, catalyzing the growth of purely rocky planets interior to the ice-rich ones~\citep{raymond18b}. Planets migrate into long chains of mean motion resonances, with the innermost planet at the inner edge of the disk. The vast majority (90-95\%) of resonant chains become unstable when the gas disk dissipates. The resulting planets match the distributions of known super-Earths~\citep{izidoro17,izidoro18b}. Given various loss process~\citep[e.g.][]{grimm93,genda05,marcus10,monteux18} the water/ice contents of these planets may be drastically overestimated. {\bf Right:} Evolution of the planet-planet scattering model for the origin of giant exoplanets~\citep[e.g.][]{adams03,chatterjee08,raymond10}. Several embryos grow quickly enough to accrete gas and grow into gas giants.  They subsequently migrate into a resonant chain without drastically affecting the orbits of nearby growing rocky planets (or outer planetesimal disks). After the disk dissipates, the vast majority (75-90\%) of giant planets systems become unstable. The resulting systems match the correlated mass-eccentricity distribution of known giant exoplanets~\citep[e.g.][]{ford08,wright09}. }  
 \label{fig:exo}
 \end{figure*}

Recent work has shown that gas accretion is far more complex than previously assumed. For few Earth-mass planets, currents of gas often pass within a small fraction of the planet's Hill sphere before exiting~\citep{fung15,lambrechts17}, casting doubts on the simple picture that gas within a planet's Hill sphere must simply cool sufficiently to approach the planet's surface.  Nonetheless, growing super-Earths must accrete gas during the disk phase~\citep{lee14,inamdar15} but should rarely reach the 50\% gas-by-mass threshold for runaway gas accretion~\citep{pollack96} because the occurrence rate of hot Jupiters is more than an order of magnitude smaller than that of super-Earths~\citep{howard10,mayor11}.  Super-Earths may be in a constant state of gas accretion~\citep{ikoma00,lambrechts17} moderated by loss processes related to collisions~\citep{schlichting15,inamdar16} as well as the dissipation of the disk itself~\citep{ikoma12,ginzburg16}.  Models invoking atmospheric loss from a few large impacts -- such as those characteristic of late instabilities in the breaking the chains model -- can broadly match the observed distribution of gaseous envelope masses~\citep{inamdar16}, although the initial, pre-impact atmospheric masses remain uncertain.

It is difficult to understand why close-in super-Earths exist in so many systems but not in all systems. Models of super-Earth formation struggle {\it not} to form them whereas microlensing observations show that similar-mass planets are extremely abundant past the snow line~\citep{beaulieu06,gould10,suzuki16b}. A simple explanation is to invoke a timing constraint: if large embryos form too slowly then they would not have time to migrate all the way to the inner edge of the disk.  However, most studies find that migration is {\it fast}~\citep[although new simulations by ][find that migration may be slower]{mcnally18} so this requires a fine-tuned delay such that most super-Earths grow large just before the gas disk dissipates. And late accretion of smaller embryos beyond a few AU after the dissipation of the disk would be quite inefficient~\citep[e.g.][]{levison01,thommes03}.

Lower disk masses also cannot explain why many stars do not host close-in super-Earths. There is an observed super-linear correlation between the disk mass and stellar mass~\citep{pascucci16}, albeit with large scatter at a given stellar mass~\citep[e.g.,][]{scholz06,williams11}. This means that M dwarf stars -- with masses between roughly 8\% and 60\% of the Sun's -- have on average significantly lower-mass disks than Sun-like stars. However, the occurrence rate of super-Earths around M dwarfs is at least as high as around Sun-like stars~\citep{howard12,fressin13,dressing15} and systems of close-in low-mass planets contain on average a higher total mass around M dwarfs~\citep{mulders15,mulders15b}.  If lower-mass disks could not produce super-Earths, M dwarfs should naively have lower average occurrence rates.

Wide-orbit giant planets may potentially explain why some systems do not have close-in super-Earths.  Once a planet accretes enough gas it carves an annular gap in the disk, slows its migration and becomes a true gas giants~\citep{lin86,bryden99,crida06}. The inward migration of more distant large embryos are then blocked by the gas giant~\citep{izidoro15a}. In this context, Uranus and Neptune (and perhaps Saturn's core) may represent failed super-Earths, embryos whose migration was blocked by the young Jupiter~\citep{izidoro15b}. This predicts an anti-correlation between systems with many close-in super-Earths and those with wide-orbit gas giants~\citep{izidoro15a}.  However, the occurrence rate of wide-orbit gas giants is $\sim$10\% for Sun-like stars~\citep{cumming08,mayor11,wittenmyer16,rowan16} and is likely far lower for M dwarfs~\citep{johnson07,lovis07,dressing15}.  For FGK stars this is a factor of 3-5 lower than the occurrence rate of super-Earths, and the problem is even worse for M dwarfs. Thus, the Jupiter migration barrier does not appear capable of explaining why {\it most} systems do not have close-in super-Earths.

Wide-orbit planets with masses comparable to the ice giants' ($10-20 \mearth$) may help solve this problem by stunting the growth of planetary embryos.  The largest planetesimals are thought to represent the seeds of planetary embryos and to grow by accreting pebbles that drift inward through the disk~\citep[e.g.,][see \S 2]{johansen17}. Once an embryo reaches the {\em pebble isolation mass} it creates a pressure bump in the gas disk exterior to its orbit that acts to trap drifting pebbles and shut off the pebble flux~\citep{morby12,lambrechts14b,bitsch18}. This not only starves the embryo but all other embryos interior to its orbit, which may continue to accrete planetesimals (but not pebbles). For typical disk parameters the pebble isolation mass is on the order of $20 \mearth$~\citep{lambrechts14b,bitsch18}. A fast-growing wide-orbit planet with a mass similar to Neptune's ($17 \mearth$) may starve the inner disk and prevent closer-in embryos from reaching large enough masses to undergo rapid migration and following the ``breaking the chains'' pattern discussed above. This mechanism is especially promising given that the abundance of wide-orbit ice giant-mass planets inferred from microlensing is on the same order as the occurrence rate of close-in super-Earths~\citep[e.g.,][]{gould10,petigura13,clanton16,winn15}.  However, given that pebble-blocking outer Neptunes must form quickly (to starve inner embryos), it is not clear how such planets could avoid migrating inward and becoming super-Earths themselves. 

It is of course possible that migration may not always follow the pattern that we have laid out. In one type of disk model that invokes winds as an angular transport mechanism, the surface density in the inner 1-2 AU of the disk increases steeply with radius~\citep{suzuki16}. In such a disk, type I migration is significantly slowed and may even be quenched~\citep{ogihara15b}, such that co-migrating super-Earths may not always form resonant chains~\citep{ogihara18}. Magnetic stresses in the midplane of certain disk models may generate positive torques that drive outward planet migration but only in specific situations, for example if the star's spin vector is aligned with the magnetic field~\citep{mcnally17,mcnally18}.  Such effects could in principle prevent inward migration in a subset of disks.

To conclude this subsection, we reiterate that the {\em breaking the chains} scenario~\citep{izidoro17,izidoro18b} provides a match to the observed super-Earth systems. The late evolution of that model should hold whether the bulk of super-Earths' mass comes from pebbles that drifted inward~\citep[the {\em drift} model; ][]{chatterjee14} or from cores that formed past the snow line~\citep[the {\em migration} model;][]{terquem07}. What remains unexplained is why so many -- but not all -- stars have close-in super-Earths.

\bigskip
\noindent
\textbf{3.2 Systems with giant exoplanets}
\bigskip

Giant exoplanets are found around $\sim$10\% of Sun-like stars~\citep[see discussion in \S 1.2; ][]{cumming08,mayor11,foremanmackey16}. Most giant exoplanets have orbital radii larger than 0.5-1 AU~\citep{butler06,udry07b} and only $\sim$1\% of stars have hot Jupiters~\citep{howard10,wright12}. Giant planet masses follow a roughly $dN/dM \propto M^{-1.1}$ distribution of minimum masses~\citep{butler06}. The median eccentricity of this population is $\sim$~0.25, roughly five times larger than Saturn and Jupiter's long-term average eccentricities~\citep{quinn91}. More massive giant planets have statistically higher eccentricities than lower-mass giant planets~\citep[typically, the division between high-mass and low-mass giant planets is at roughly Jupiter's mass;][]{jones06,ribas07,ford08,wright09}. 

As for super-Earths, the population of giant exoplanets is thought to have been sculpted in large part by migration and instability.  

A subset of giant exoplanets has been found to be in mean-motion resonance. Strong resonances with small-amplitude libration of resonant angles are thought to be a clear signature of migration (e.g., \cite{papaloizou06,kley12}; weaker resonances can be produced by instabilities -- see \cite{raymond08b}). Notable examples of giant exoplanets in resonance include the GJ 876 system in which three planets are locked in 4:2:1 Laplace resonance~\citep{rivera10} and the HR 8799 system~\citep{marois08,marois10}, for which stability considerations indicate that the outer three or perhaps all four super Jupiter-mass planets may be in resonance~\citep{reidemeister09,fabrycky10,godziewski14,gotberg16}. These systems are thought to represent signature cases of migration, and a recent analysis found that a significant fraction may be in resonance~\citep[25\% of the 60 giant exoplanet systems studied in][]{boisvert18}.

Giant planets are thought to form on circular orbits. However, a large fraction of giant exoplanets are found to have significant orbital eccentricities, which are thought to be an indicator of dynamical instability~\citep[see, e.g.][]{ford08}. The eccentricity distribution of giant exoplanets can be reproduced by the {\it planet-planet scattering} model, which proposes that the observed planets are the survivors of system-wide instabilities~\citep{rasio96,weidenschilling96,lin97}. Giant planets are assumed to form in systems of two or more planets.  After the gaseous disk dissipates the planets' orbits become dynamically unstable, leading to a phase of close gravitational scattering events. Scattering events involve orbital energy and angular momentum exchange between the planets and tend to increase their eccentricities and inclinations. During this phase of dramatic orbital excitation, nearby small bodies such as the building blocks of terrestrial planets are generally driven to such high eccentricities that they collide with the host star~\citep{veras05,veras06,raymond11,raymond12,matsumura13}. More distant planetesimals are preferentially ejected~\citep{raymond11,raymond12,raymond18,marzari14}. Giant planet instabilities typically conclude with the ejection of one or more planets into interstellar space~\citep{veras12}. The surviving planets have eccentric orbits~\citep{ford03,adams03,chatterjee08,ford08,raymond10}.  The observed eccentricity distribution can be matched if at least 75\% -- and probably closer to 90\% -- of all giant exoplanet systems represent the survivors of instabilities~\citep{juric08,raymond10}.  Scattering can also match the observed orbital spacing of giant planet systems, in particular with regards to their proximity to the analyically-derived boundary for orbital stability~\citep[also called `Hill stability'; ][]{raymond09a}, as well as the secular structure of observed systems~\citep{ford05,timpe13}.  

Gravitational scattering produces an energy equipartition among planets in the same system such that the lowest-mass planets have the highest eccentricities. This is in disagreement with observations, which show that higher-mass planets have higher eccentricities than lower-mass, with a statistically significant difference~\citep{ribas07,ford08,wright09}. This discrepancy is resolved if systems with very massive giant planets ($M \gtrsim M_{Jup}$) systematically form multiple, very massive giant planets with near-equal masses~\citep{raymond10,ida13}. The eccentricities of surviving planets are highest in systems with the most massive, equal-mass planets~\citep{ford01,raymond10}.  

All that is needed to trigger instability is the formation and migration of 2-3 gas giants\footnote{Wide binary stars can also trigger instabilities, as torques from passing stars and the galactic tidal field occasionally shrink their pericenters to approach the planetary region~\citep{kaib13}.}. The timescale of instability is a function of the planets' initial separations~\citep{chambers96,marzari02}, so most studies simply started planets in unstable configurations to determine the outcome of the instability.  A more self-consistent approach invokes a prior phase of orbital migration. While migration is often thought of as a dynamically calm process, several studies have shown that migration of multiple giant planets often generates instabilities after, or even during, the gaseous disk phase~\citep{moeckel08,matsumura10,marzari10,moeckel12,lega13}. The planets that emerge from migration-triggered instability provide a match to the observed giant exoplanets~\citep{adams03,moorhead05}. 

Hot Jupiters present an interesting melding of migration and instability~\citep[see][for a review]{dawson18}. In recent years it has been debated whether hot Jupiters migrated in to their current locations~\citep{lin96,armitage07} or were scattered to such high eccentricities (and such small pericenter distances) that tidal dissipation within the star (also called `tidal friction') shrank their orbits~\citep{nagasawa08,beauge12}. The Kozai effect -- in which a perturbing planet or star on a highly-inclined orbit induces large-scale, anti-correlated oscillations in a planet's eccentricity and inclination -- may play a role in generating the high eccentricities needed to produce hot Jupiters by tidal friction~\citep{fabrycky07,naoz11} but only in situations in which such large mutual inclinations arise (e.g., after an instability). Of course, it is possible that hot Jupiters underwent rapid gas accretion close-in and thus represent the rare hot super-Earths that grew fast enough to trigger runaway gas accretion~\citep{bodenheimer00,boley16,batygin16b}. Yet even if hot Jupiters accreted their gas close-in, the arguments presented in \S 3.1 still indicate that their building blocks must have either migrated or drifted inward. 

Another constraint comes from observations of the Rossiter-McLaughlin effect, which measures the projected stellar obliquity in systems with transiting planets~\citep{winn05,gaudi07}. This translates to a projection of the planet's orbital inclination with respect to the stellar equator. For a significant fraction of measured hot Jupiters, the orbital plane is measured to be strongly misaligned with the host stars' equators. More massive stars, for which the timescale for tidal dissipation are much longer~\citep{zahn77}, are far more likely to host misaligned hot Jupiters~\citep{winn10,triaud10,albrecht12}. This may be explained by planets being scattered onto very eccentric and inclined orbits before their orbits are shrunk by tidal friction~\citep{fabrycky07,nagasawa08,naoz11,beauge12,lai12}. 

Migration predicts that hot Jupiters should remain aligned with their birth disks.  However, disks themselves can be torqued into configurations that are misaligned with respect to the stellar equator~\citep[e.g.][]{lai11,batygin12}. If planet-forming disks are themselves misaligned then both the migration and close-in growth models can in principle explain hot Jupiters' misaligned orbits. In the Kepler-56 system two massive planets share an orbital plane that is misaligned with the stellar equator~\citep{huber13}. While this is suggestive of the planets having formed in a tilted disk, other dynamical mechanisms can plausibly explain such tilting~\citep{innanen97,mardling10,kaib11,boue14,gratia17}.  In addition, there is as yet no sign of debris disks -- dust disks observed around older stars whose gas disks have already dissipated -- that are misaligned with the equators of their host stars~\citep{greaves14}.

While the general picture of migration and instability appears to match the broad characteristics of giant exoplanets, questions remain. While slower than for low-mass planets, the timescale for type 2 migration -- which is thought to be controlled in large part by the disk's viscosity~\citep{lin86,ward97,durmann15} -- is still in many cases faster than the disk lifetime. Why, then are there so few gas giants interior to 0.5-1 AU?  Photo-evaporation of the disk produces inner cavities of roughly that size~\citep{alexander14,ercolano17}. The cavity is only generated late in the disk's lifetime, so if it is to explain the deficit of gas giants within 0.5-1 AU this would require very slow migration and thus very low-viscosity disks (\cite{alexander12b}; but see \cite{wise18} and discussion in \cite{morbyraymond16}).  


To conclude, the general evolution of gas giant systems thus appears to follow a similar pattern as super-Earths' {\em breaking the chains} evolution. Gas giants form in cohorts and migrate into resonant configurations~\citep[e.g.,][]{kley12}. After the disk dissipates (or sometimes before) the vast majority of systems become unstable and undergo a violent phase of planet-planet scattering that generates the observed gas giant eccentricities~\citep[e.g.,][]{juric08,raymond10} and also disrupts the growth of any smaller planets in the systems~\citep[e.g.,][]{raymond11}. The origins of hot Jupiters remain debated, but viable formation models invoke a combination of migration and instability~\citep[see][]{dawson18}.

\section{\textbf{MODELS FOR SOLAR SYSTEM FORMATION}}

We now turn our attention to the origin of the Solar System. Given the astrobiological context of this chapter, we emphasize the origin of the inner Solar System. However, it is important to keep in mind that dynamical perturbations from Jupiter during its growth, migration and early evolution played an important part in shaping the terrestrial planets. 

In this section we first lay out the observational constraints (\S 4.1).  Next, in \S 4.2 we present a rough timeline of events, including a discussion of the late instability in the giant planets' orbits (the so-called 'Nice model'). In \S 4.3 we describe the so-called {\em classical model} of terrestrial planet formation and explain its shortcomings. In \S 4.4 we present and contrast three models for the early evolution of the inner Solar System. In \S 4.5 we explore a feature of the Solar System that remains hard to explain with all current models: the mass deficit in the very close-in Solar System.

\bigskip
\noindent
\textbf{4.1 Solar System constraints}
\bigskip

A successful formation model must match the Solar System's broad characteristics (see \S 7.3 for a philosophical discussion). We now lay out the central constraints to be matched, in rough order of importance.  We start with constraints related to the inner Solar System (roughly from most- to least- stringent) and conclude with constraints related to the outer Solar System (Jupiter and beyond). 

{\bf The masses and orbits of the terrestrial planets.}  The terrestrial planets follow an odd pattern, with two large central planets (Venus and Earth) flanked by much smaller ones (Mercury and Mars). Roughly 90\% of all the rocky material in the Solar System is thus concentrated in a ring that is only 0.3 AU in width (encompassing the orbits of Venus and Earth). In addition, the terrestrial planets' orbits are remarkably close to circular. These constraints are often quantified using statistics on the radial mass concentration~\citep{chambers01} and degree of orbital excitation~\citep[often using the {\em angular momentum deficit}, defined as the fractional deficit in angular momentum of a system of planets relative to an identical system on perfectly circular, coplanar orbits; ][]{laskar97,chambers01}. In addition, the Earth/Mars and Venus/Mercury mass ratios offer simple, surprisingly strong constraints.

{\bf The mass, orbital and compositional structure of the asteroid belt.} The main belt (between roughly 2 and 3.2 AU) covers a surface area more than three times larger than that of the terrestrial planet region, yet the entire belt contains only $\sim 4.5 \times 10^{-4} \mearth$~\citep{krasinsky02,kuchynka13,demeo13}. Yet the asteroids' orbits are excited, with eccentricities that span from zero to 0.3 and inclinations that extend above $20^\circ$. The belt contains a diversity of spectroscopically-distinct types of objects~\citep{bus02}. The belt is also radially segregated: the inner main belt (interior to $\sim 2.7$~AU) is dominated by S-types whereas the main belt beyond 2.7 AU is dominated by C-types~\citep{gradie82,demeo13,demeo14}. S-types are identified with ordinary chondrites, which are relatively dry (with water contents below 0.1\% by mass), whereas C-types are associated with carbonaceous chondrites, which typically contain $\sim 10\%$ water by mass~\citep{robert77,kerridge85,alexander18}.

{\bf The cosmochemically-constrained growth histories of Earth and Mars.}  Isotopic analyses of different types of Earth rocks, lunar- and Martian meteorites constrain the growth histories of Earth and Mars. Isotopic systems such as Hf-W with half-lives comparable to the planets' formation timescales are particularly useful~\citep[the half-life of radioactive $^{182}$Hf is 9 Myr; see][]{alexander01}.  These studies indicate that Earth's core formation did not finish until at least $\sim 40-100$ million years after the start of planet formation~\citep{touboul07,kleine09}. The final episode of core formation on Earth is generally assumed to have been the Moon-forming impact~\citep{benz86,canup01}. Mars' growth is directly constrained to have been far faster than Earth's~\citep{nimmo07}. Indeed, Mars' accretion was complete within 5-10 Myr~\citep{dauphas11}. 

{\bf The abundance and isotopic signature of water on Earth.} Despite being mostly dry and rocky, Earth still contains a small fraction of water by mass, and is thought to be essential for life. The exact amount of water on Earth remains only modestly-well constrained. An ``ocean'' of water is defined as the total amount of water on Earth's surface, roughly $1.5 \times 10^{24}$ grams (or $\sim$0.025\% of an Earth mass). The mantle is thought to contain between a few tenths of an ocean~\citep{hirschmann06,panero17} and 5-10 oceans~\citep{lecuyer98,marty12,halliday13}. The core is generally thought to be very dry~\citep{badro14} but \cite{nomura14} inferred a very large reservoir of water exceeding 50 oceans. Assuming a total water budget of four oceans, Earth's bulk water content is thus 0.1\% by mass.  

The isotopic signature of Earth's water -- the D/H and $^{15}$N/$^{14}$N ratios -- is a key discriminant of different models of water delivery (see \S 6).  Earth's water is a good match to carbonaceous chondrite meteorites, specifically the CM subgroup~\citep{marty06,alexander12}. Earth's water is isotopically distinct from nebular and cometary sources~\citep[see data compiled in ][; note that there are two comets observed to have Earth-like D/H ratios but both have non-Earth-like $^{15}$N/$^{14}$N ratios; see discussion in \S 6]{morby00}.

{\bf The late veneer on Earth, Mars and the Moon.} Highly-siderophile elements are those that are thought to have a chemical affinity for iron rather than silicates. Most of these elements are thus thought to be sequestered in a planet's core during core-mantle segregation. All of the highly-siderophile elements in Earth's crust must therefore have been delivered by impacts after the Moon-forming impact~\citep{kimura74}. From the abundance of highly-siderophile elements, and assuming the impactors to be chondritic in compositions, it has been inferred that the last $\sim 0.5\%$ of Earth's accretion took place after the Moon-forming impact~\citep{day07,walker09,morbywood15}.  Meteorite constraints indicate that Mars accreted $\sim 9$ times less material than Earth during the late veneer, and that the Moon accreted 200-1200 times less than Earth~\citep{day07,walker09}.  

{\bf The orbits and masses of the giant planets.}  The giant planets are radially segregated by mass, with the most massive planets closest-in. As discussed in \S 1.1, Jupiter's orbit is wider than most known giant exoplanets' and it is only barely detectable by long-duration radial velocity surveys. Jupiter and Saturn each have low-eccentricity but non-circular orbits, each with Myr-averaged eccentricities of $\sim 0.05$~\citep{quinn91}. Uranus' average eccentricity is comparable to the gas giants' but Neptune's is only $\sim 0.01$. There are no mean motion resonances among the giant planets. The Jupiter/Saturn, Saturn/Uranus and Uranus/Neptune period ratios are 2.48, 2.85, and 1.95, respectively. 

{\bf The total mass and orbital structure of the outer Solar System's small body populations.}  The Kuiper belt extends outward from just Neptune's orbit. It has a complex orbital structure that includes a population of objects such as Pluto that are locked in mean motion resonances with Neptune. The Kuiper belt has a broad eccentricity and inclination distribution and includes a population of very dynamically cold objects from 42-45 AU often called the cold classical belt.  The total mass in the Kuiper belt has been estimated at a few to ten percent of an Earth mass~\citep{gladman01}. The scattered disk is a subset of Kuiper belt objects whose orbits cross those of the giant planets. The Oort cloud is the source of long-period (isotropic) comets and extends from roughly 1,000 AU out to the Sun's ionization radius~\citep[currently at $\sim 200,000$~AU; see][for details]{tremaine93}.

\bigskip
\noindent
\textbf{4.2 A rough timeline of events}
\bigskip

Theory and observations can combine to provide a rough timeline of the events that must have taken place in the Solar System. Time zero is generally assumed to be the time of formation of CAIs (Calcium and Aluminum-rich Inclusions), roughly mm-sized components of primitive (chondritic) meteorites that are well-dated to be 4.568 Gyr old~\citep[e.g.,][]{bouvier10}.

\begin{itemize}
\item Within 100,000 years planetesimal formation was underway. CAIs and mm-scale chondrules had started to form~\citep[e.g.,][]{connelly08,nyquist09} and coalesce into larger objects~\citep[e.g.][]{dauphas11b,johansen15}.\footnote{The origin of chondrules is hotly debated, and some models suggest that they are the outcomes of collisions between planetesimals and planetary embryos rather than their building blocks~\citep[e.g.][]{asphaug11,johnson15,lichtenberg18}. } 

\item Within 1 million years large embryos had formed. Ages of iron meteorites indicate that embryos had formed in the inner Solar System~\citep[e.g.,][]{halliday06,kruijer14,schiller15}. Meanwhile, the segregation of the parent bodies of carbonaceous and non-carbonaceous meteorites indicates that at least one $\sim 10 \mearth$ embryo -- presumably Jupiter's core~\citep{kruijer17} -- had formed in the giant planet region. From this point onward, this core blocked the inward drift of pebbles and thus starved the inner Solar System~\citep[e.g.,][]{bitsch18}.

\item Within a few million years the gaseous planet-forming disk had dissipated.  Evidence for the timescale of disk dissipation comes from two sources.  First, observations of hot dust -- thought to trace the gas -- around stars in young clusters with different ages indicate a typical dissipation timescale of 2-5 Myr~\citep{haisch01,hillenbrand08b,pascucci09,mamajek09}.  Second, given that all chondrule formation models require the presence of the gas disk, the latest-forming chondrules provide a lower limit on the gas disk's lifetime of 4-5 Myr~\citep[the CB chondrites; see ][]{kita05,krot05,johnson16}. The existence of a hot Jupiter around a 2 million-year-old T Tauri star~\citep{donati16} demonstrates that giant planet formation and migration can happen on an even shorter timescale. The gas- and ice giant planets were fully-formed and likely in a compact resonant resonant chain~\citep{morby07,izidoro15b}. By this time Mars was close to fully-formed~\citep{nimmo07,dauphas11} but Earth (and presumably Venus) were still actively accreting via planetesimal and embryo impacts.

\item While the Sun was still in its birth cluster it underwent a relatively close encounter with another star.  Such an encounter has been invoked to explain the orbits of the Sednoids~\citep{morby04,kenyon04,jilkova15,pfalzner18} -- named after Sedna, the first one discovered~\citep{brown04} -- whose semimajor axes are greater than 250 AU and whose perihelia are detached from the planets'.  The encounter may have either excited existing Solar System planetesimals onto Sedna-like orbits or captured the objects from the passing star. The encounter distance was likely at a few hundred to a thousand AU.  While the exact properties of the Sun's birth cluster remain a matter of debate~\citep{adams10,gounelle12,portegies18}, such encounters are expected to be a common occurrence~\citep{malmberg11}. It is interesting to note that the Sun must have left its parent cluster before the giant planet instability, since that is when the Oort cloud would have formed~\citep{brasser13} and it would be much more compact had it formed in a cluster environment, given the stronger tidal field~\citep{tremaine93,kaib08}. 

\item Roughly 50-100 Myr after CAIs, Earth suffered its final giant impact~\citep{touboul07,kleine09}. This impact triggered Earth's final core formation event and the formation of the Moon~\citep{benz86,canup01}. Only $\sim0.5\%$ of Earth's mass was accreted after this point~\citep{day07,walker09,morbywood15}. 

\item Within 500 Myr the outer Solar System went unstable. The instability -- thought to have been generated by interactions between the giant planets and an outer planetesimal disk, essentially the primordial Kuiper belt -- is commonly referred to as the {\em Nice model}. Starting from a compact resonant chain originally formed as a consequence of a previous phase of migration in the gaseous disk~\citep{morby07}, the giant planets underwent a series of close encounters. Interactions with the outer planetesimal disk caused the giant planets' orbits to radially spread out and destabilized the planetesimal disk~\citep{levison11}, which led to a phase of impacts throughout the Solar System that was originally proposed to correspond to the so-called late heavy bombardment~\citep{tera74,gomes05}, the event often associated with many of the oldest craters on Mercury, the Moon and Mars. The instability can explain a number of features of the Solar System including the giant planets' present-day orbits~\citep{tsiganis05,nesvorny12}, the orbital distribution of Jupiter's co-orbital asteroids~\citep[in particular their large inclinations][]{morby05,nesvorny13}, and the characteristics of the giant planets' irregular satellites~\citep{nesvorny07}. Simulations that invoke that the young Solar System had 1-2 additional ice giants that were ejected during the instability have a much higher success rate in matching the present-day Solar System~\citep{nesvorny12,batygin12b}. The gas giants' relatively low eccentricities constitute a dynamical constraint: Jupiter and Saturn never underwent a close mutual encounter, although they must have had encounters with one or more ice giants~\citep{morby09b}.

While the existence of the instability remains in favor, the late timing has recently been challenged~\citep{boehnke16,morby18,michael18,nesvorny18}. An early giant planet instability is easier to understand from a dynamical perspective. The dispersal of the gaseous disk is the natural trigger for instabilities~\citep[e.g.,][]{matsumura10} and simulations had substantial difficulty in delaying the onset of instability~\citep{gomes05,levison11}. Simulations of giant planets interacting with outer planetesimal disks indeed show that most instabilities happen early, although there is a tail of instabilities that extends to much longer timescales~\citep{thommes08b,raymond10}. A key input in such models -- the inner edge location and orbital distribution of planetesimals in the outer primordial disk -- remains poorly-constrained.  

The giant planet instability must have had a significant impact on the inner Solar System. The changing dynamical environment caused by changes in Jupiter and Saturn's orbits caused secular resonances to sweep and/or jump across the inner Solar System, exciting anything in their path~\citep{brasser09,agnor12}. A late instability tends to excite and often to destabilize the orbits of the already-formed terrestrial planets~\citep{brasser13,roig16,kaib16}. An early instability -- triggered shortly after the gas disk's dispersal and before the final assembly of the terrestrial planets -- has the potential to resolve this problem, and constitutes the basis for one of the terrestrial planet formation models we will discuss in \S 4.4~\citep[the {\em Early Instability} model of][]{clement18}. 

\item For the past 4 billion years, the orbital architecture of the Solar System has remained roughly constant. Most impacts on the terrestrial planets come from asteroids that are disrupted and whose fragments end in unstable resonances~\citep[often after drifting due to the Yarkovsky effect; see][]{gladman97,bottke06b,granvik17}. The planets' orbits undergo secular oscillations due to long-range gravitational perturbations~\citep[e.g.][]{quinn91}. The oscillations in Earth's orbit and spin are called Milankovitch cycles and play a key role in its climate evolution~\citep[e.g.,][]{berger05}. The inner Solar System is chaotic with a Lyapunov timescale of a few Myr~\citep{laskar90,batygin15}, but it is unknown whether the outer Solar System's evolution is chaotic or regular, as both types of solutions exist within the current error bars on the giant planets' positions~\citep{hayes07}. Regardless of whether the outer Solar System is chaotic or regular there is no chance of future instability.  In contrast, the terrestrial planets have a $\sim 2\%$ chance of becoming unstable before the Sun becomes a red giant in 4-5 Gyr~\citep{laskar09}.
\end{itemize}

\bigskip
\noindent
\textbf{4.3 The classical model of terrestrial planet formation}
\bigskip

The so-called {\em classical} model of terrestrial planet formation was pioneered by a series of papers by George Wetherill spanning 2-3 decades~\citep[e.g.,][]{wetherill78,wetherill85,wetherill96}.  It has succeeded in explaining a large number of features of the inner Solar System, and its shortcomings have served to point newer models in the right direction. The classical model remains to this day the basis of comparison with more recent models~\citep[e.g.][]{morby12b,raymond14,jacobson15,izidoro18}.

The central assumption in the classical model is that giant planet formation can be considered separately from terrestrial accretion. At face value this appears to be a reasonable assumption. Gas-dominated protoplanetary disks are observed to dissipate in a few Myr~\citep{haisch01,hillenbrand08b}, setting an upper limit on the timescale of gas giant formation.  In contrast, cosmochemical studies have demonstrated that Earth's accretion lasted 50-100 Myr~\citep{kleine09}. Simulations of the classical model start from a population of rocky building blocks (planetary embryos and planetesimals) and fully-formed gas giants, inherently assuming that there was no prior interaction between these different populations and that the gas had already been removed.

\begin{figure}[t]
 \epsscale{0.95}
 \plotone{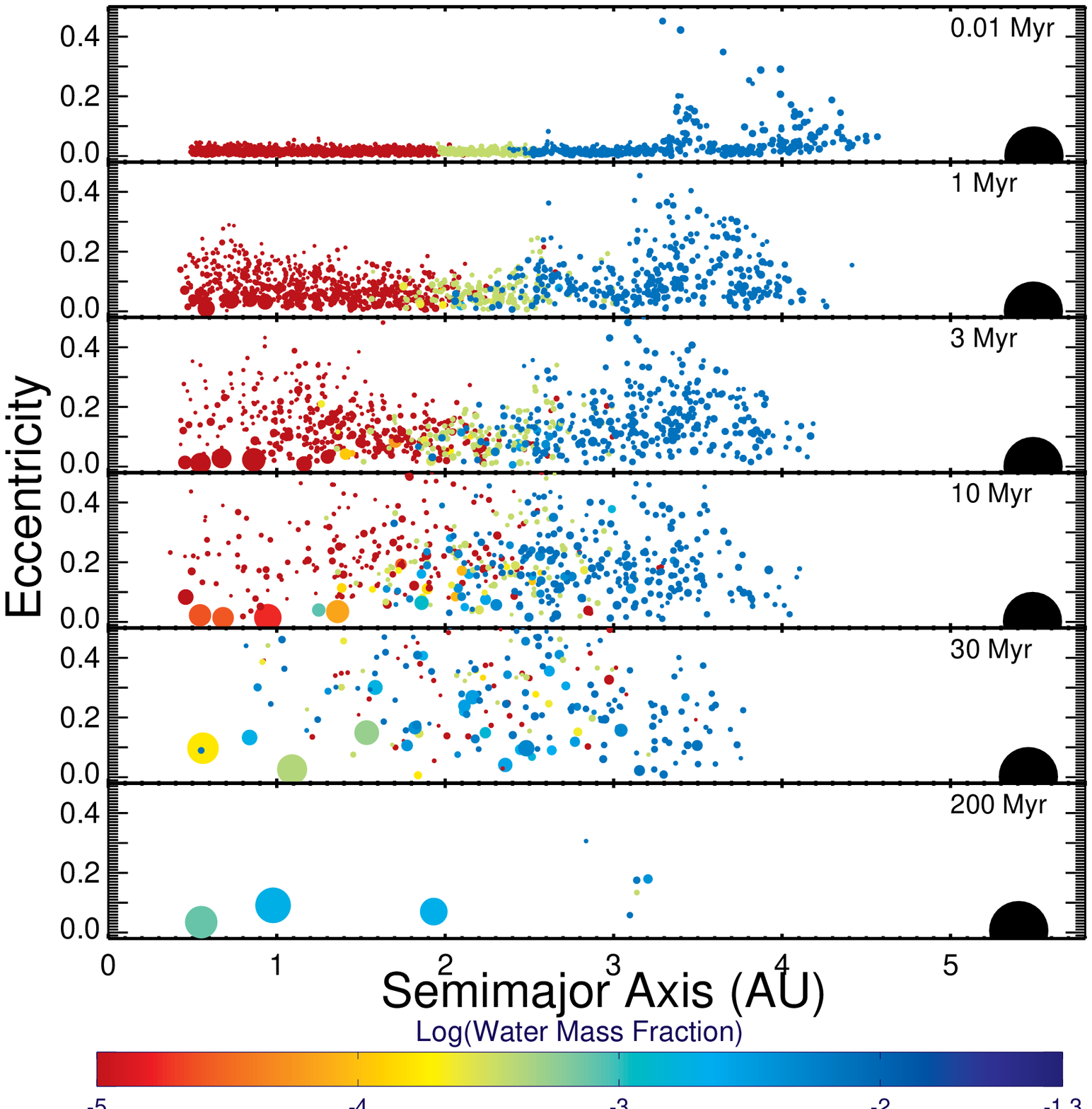}
 \caption{\small A simulation of the classical model of terrestrial planet formation~\citep[adapted from ][]{raymond06b}. The simulation started from 1886 self-gravitating planetary embryos, represented as dots with size proportional to its mass$^{1/3}$. Jupiter is fully-formed (large black dot) on a near-circular orbit at the start of the simulation. The color of each embryo represents its water content (see color bar at the bottom), with red objects being dry and the darkest blue containing 5\% water by mass~\citep[see][]{raymond04}.  This simulation produced quite good Earth and Venus analogs, a very poor Mars analog, and a plausible, albeit far too massive, asteroid belt. Note that time zero for this simulation corresponds to the dissipation of the gaseous disk, a few Myr after CAIs. A movie of this simulation can be viewed here: https://youtu.be/m7hNIg9Gxvo.}  
 \label{fig:classical}
 \end{figure}

Figure~\ref{fig:classical} shows the evolution of a characteristic simulation of the classical model~\citep[from][]{raymond06b}. The simulations is gas-free and so its time zero effectively corresponds to the dissipation of the gaseous disk. The population of rocky embryos (initially $\sim$Ceres- to Moon-mass in this case) self-excites by mutual gravitational forcing from its inner regions outward, producing larger embryos with a characteristic spacing~\citep{kokubo98,kokubo00,kokubo02}.  The outer parts of the rocky disk are excited by secular and resonant forcing from Jupiter, and excited bodies transmit this disturbance through mutual gravitational scattering.  There is a long chaotic phase characterized by excitation of planetesimals to high eccentricities; the embryos' eccentricities and inclinations are generally kept lower by dynamical friction~\citep{obrien06,raymond06b}. During this phase embryos grow by accreting planetesimals as well as other embryos, and given that other embryos are growing concurrently, the largest impacts tend to happen late~\citep[e.g.][]{agnor99}. By roughly 100 Myr after the start of the simulation most remnant planetesimals have been cleared out and three planets have formed.  

The simulation from Fig.~\ref{fig:classical} illustrates the successes of the classical model as well as its shortcomings.  The two inner surviving planets bear a strong likeness to Venus and Earth.  Their orbital separation and eccentricities are similar and their masses are reasonably close.  In addition, their feeding zones are wide enough to extend into the outer asteroid belt and have accreted water-rich material~\citep[see][]{morby00}. Earth's accretion happened on a geochemically-appropriate timescale of $\sim 100$~Myr and included late giant impacts suitable for Moon formation. However, the third planet bears little resemblance to Mars.  Its orbit is somewhat wide of Mars' but the big problem is that the planet is as massive as Earth.  

Mars is the classical model's Achilles heel. Simulations of the classical model systematically fail to match Mars' small mass and instead form Mars analogs that are a factor of 5-10 too massive~\citep{wetherill78,chambers01,raymond06b,raymond09c,morishima10,fischer14b,kaib15}. This was first pointed out by \cite{wetherill91} and is commonly referred to as the `small Mars' problem.  

The small Mars problem can be understood in a very simple way. If we assume that the disk of rocky building blocks extended smoothly from within 1 AU out to the giant planet region, then there was roughly the same amount of mass in Mars' feeding zone as in Earth's. In the absence of large perturbations, bottom-up accretion therefore produces Mars analogs that are as massive as Earth.

There are some circumstances under which the classical model can produce small Mars analogs. For example, if Jupiter and Saturn's orbits were more excited ($e_{Jup} \approx e_{Sat} \approx 0.07-0.1$) during terrestrial accretion than they are today, then secular resonances would have been far stronger and could have acted to clear material from the Mars zone without depleting Earth's feeding zone~\citep[the {\em EEJS}, or `Extra Eccentric Jupiter and Saturn' configuration from][]{raymond09c,morishima10,kaib15}. However, the EEJS setup has its own Achilles heel: its initial conditions are not consistent with the evolution of Jupiter and Saturn in the gaseous disk. Simulations universally show that planet-disk interactions tend to drive the planets into resonance~\citep[in this case, specifically the 3:2 or 2:1 resonances][]{pierens14}. However, if Jupiter and Saturn were in a resonant configuration, the location of their secular resonances within the terrestrial disk would not help to produce a small Mars~\citep[e.g.][]{izidoro16}. A similar model invokes secular resonance sweeping during the dispersal of the gaseous disk to explain the depletion of the asteroid belt and Mars region~\citep{nagasawa05,thommes08c,bromley17}.  However, this model suffers from the same problem as the EEJS model: the gas giants' orbits are not consistent with the evolution of the disk, and using appropriate (generally lower-eccentricity, resonant) orbits removes the desired depletion. An early giant planet instability may, however, produce a giant planet configuration similar to the EEJS configuration as we discuss in \S 4.4.

The small Mars problem is inherently coupled with the asteroid belt's orbital excitation~\citep{izidoro15c}. While very low in total mass, the asteroids' orbits are much more excited than the planets', with a broad range of eccentricities and inclinations. The current amount of mass in the belt cannot account for its excitation because there is not enough mass for gravitational self-stirring to be efficient~\citep{morby15b,izidoro16}. Yet a depleted region extending from Earth to the belt may explain why Mars is so small~\citep{izidoro14b}. Indeed, the terrestrial planets' orbits are well-matched if they formed from a narrow ring of embryos that only extended from 0.7-1 AU~\cite{hansen09,walsh16,raymond17b}. At face value, this means that a low-mass Mars implies an underexcited asteroid belt, and an appropriately excited asteroid belt implies a Mars that is far too massive~\citep{izidoro15c}.  

The small Mars and asteroid excitation problems are the primary shortcomings of the classical model.  However, the model also cannot account for Mercury's small mass relative to Venus, although this remains a struggle for all models~\citep[see, e.g., ][]{lykawka17}.

\bigskip
\noindent
\textbf{4.4 Viable models for the inner Solar System}
\bigskip

\begin{figure*}
 \epsscale{1.95}
 \plotone{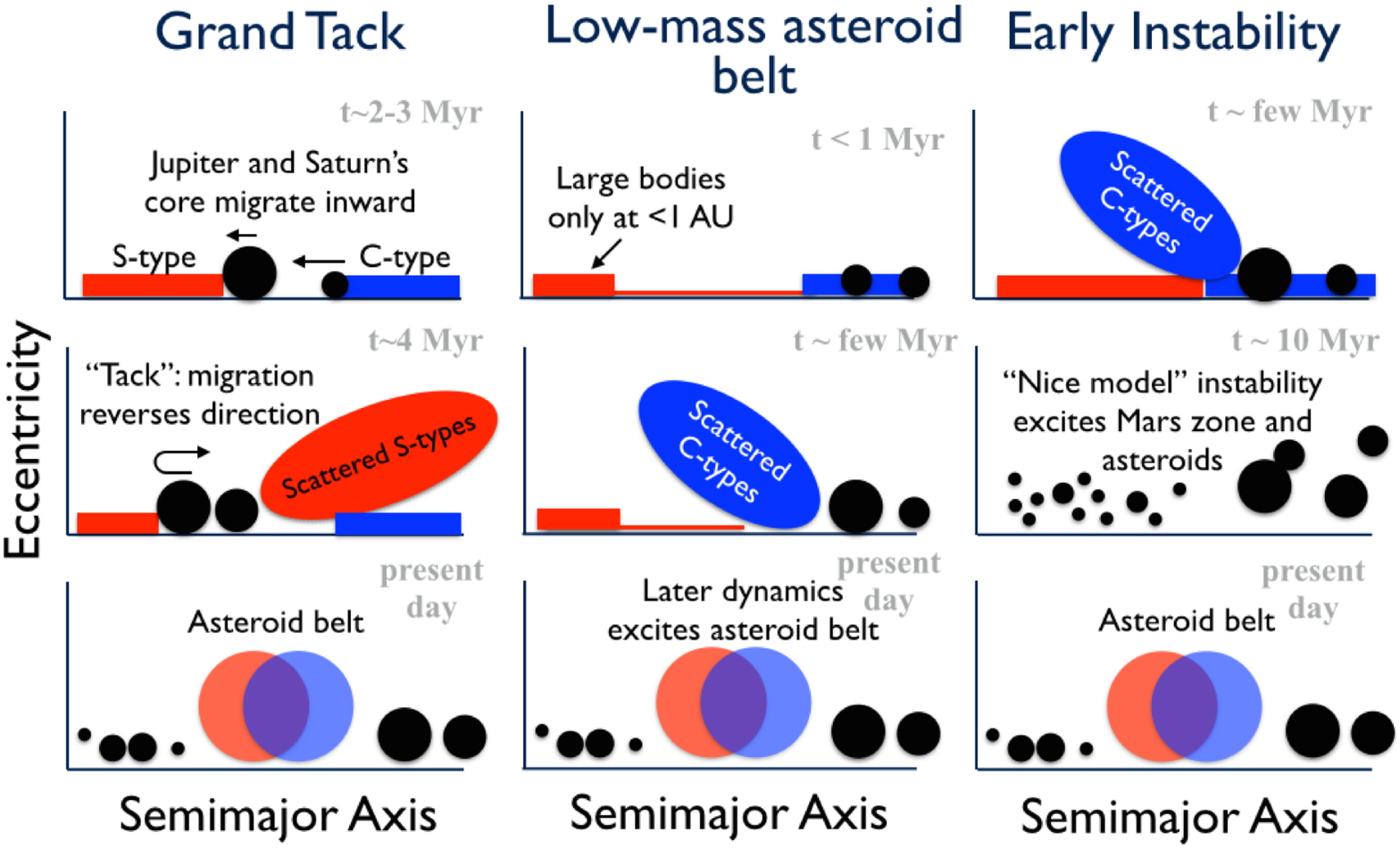}
 \caption{\small Illustration of three models that can each match the large-scale properties of the inner Solar System. {\bf Left:} The Grand Tack model~\citep{walsh11,raymond14c,brasser16}, invokes Jupiter's inward-then-outward migration to truncate the inner disk of rocky material, to explain the large Earth/Mars mass ratio. The asteroid belt is emptied then re-populated from two different source regions. S-types from interior to Jupiter's initial orbit were scattered outward during Jupiter's inward migration then back inward during its outward migration and implanted back onto similar orbits to their original ones, albeit with a low efficiency. C-types were implanted from exterior to Jupiter's initial orbit, also during the outward migration phase. {\bf Center:} The Low-mass Asteroid Belt model~\citep{hansen09,drazkowska16,raymond17b}, which proposes that the bulk of rocky planetesimals in the inner Solar System formed within a narrow ring that only extended from roughly 0.7 to 1 AU. The ring produced terrestrial planets with the correct properties. Meanwhile, the asteroid belt was contaminated with C-type planetesimals implanted during the giant planets' growth~\citep{raymond17}.  Later dynamical evolution -- perhaps explained by chaotic evolution of Jupiter and Saturn~\citep{izidoro16} or by secular forcing during the giant planet instability~\citep{deienno18} -- is required to explain the asteroids' level of excitation. {\bf Right:} The Early Instability model~\citep{clement18}, which is based on the giant planet instability happening within $\sim 10$~Myr after the dispersal of the gaseous disk. The instability acts to excite and deplete the asteroid belt and Mars' feeding zone, leading to realistic Earth/Mars mass ratios and terrestrial planets. Simulations find that the terrestrial system is best matched when the surviving giant planets match the real ones. }  
 \label{fig:comparison}
 \end{figure*}

We now discuss three successful models for the origin of the inner Solar System: the Low-mass Asteroid belt, Grand Tack, and Early Instability models (summarized in Fig.~\ref{fig:comparison}). We explain the central assumptions of each model, what circumstances are required for the key mechanisms to operate, and how to test or falsify them. We order the models by when Mars' feeding zone was depleted, from earliest  to latest.

{\bf The Low-mass Asteroid Belt model} proposes that Mars is small because simply because very few planetesimals formed between Earth's orbit and Jupiter's. Planetesimal formation has been shown to depend strongly on the gas disk's {\em local} properties~\citep[e.g.][]{simon16,yang17}. While gas disks are expected to have a relatively smooth radial distributions, ALMA observations show that dust in young disks is concentrated into rings~\citep{alma15,andrews16}. It is not at all clear that planetesimals should form uniformly across the disk.  Indeed, \cite{drazkowska16} modeled dust coagulation and drift in an evolving gas disk and found rings of planetesimals produced by the streaming instability centered at roughly 1 AU (see \S 2). Additional mechanisms such as vortices can also act to strongly concentrate particles at $\sim 1$~AU to produce planetesimal rings~\citep{surville16,surville18}.  

The Low-mass Asteroid belt model thus starts from a ring of planetesimals containing roughly $2 \mearth$ centered between Venus' and Earth's present-day orbits. The terrestrial planets that accrete from such a planetesimal annulus provide a good match to the terrestrial planets' radial mass distribution~\citep{hansen09,kaib15,walsh16,raymond17b}. In this context, Mars' growth was stunted when it was scattered out of the dense ring of embryos. This naturally explains why Mars stopped accreting early~\citep{dauphas11}. Earth's growth was more prolonged, lasting up to $\sim 100$~Myr. 

\begin{figure}[t]
 \epsscale{0.99}
 \plotone{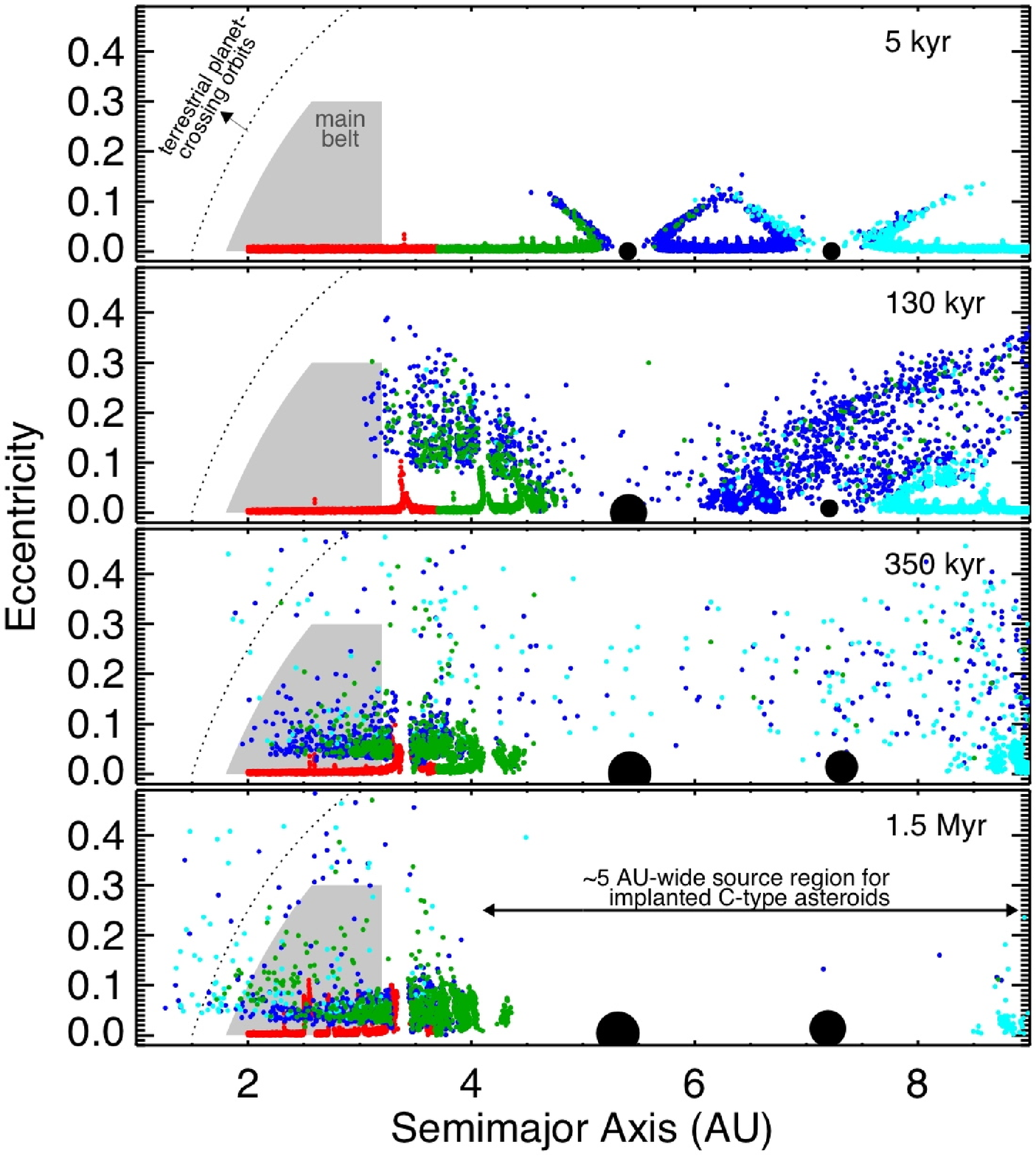}
 \caption{\small Injection of planetesimals into the asteroid belt and toward the terrestrial planet region as a consequence of giant planet formation~\citep[from][]{raymond17}. This simulation starts with Jupiter and Saturn's $3 \mearth$ cores embedded in a realistic gas disk~\citep[disk model from][adapted to include gaps carved by the planets]{morby07a} including a population of $10^4$ planetesimals assumed to be 100~km in diameter for the gas drag calculation. Jupiter grew (and carved a gap in the disk) from 100-200 kyr and Saturn from 300-400 kyr, and the disk dissipated on a 200 kyr exponential timescale (uniformly in radius). The colors of planetesimals represent their starting location. This case is the least dynamic possible scenario as it neglects migration of the gas giants and the formation and migration of the ice giants. Including those factors the source region of planetesimals implanted into the main belt extends out to $\sim 20$~AU~\citep{raymond17}. An animation of this simulation can be viewed here: https://youtu.be/Ji5ZC7CP5to.}  
 \label{fig:injection}
 \end{figure}

The compositional diversity of the asteroid belt in this scenario can be explained as a simple byproduct of the giant planets' growth~\citep[see Fig.~\ref{fig:injection}; from ][]{raymond17}. Jupiter's (and later, Saturn's) phase of rapid gas accretion invariably destabilized the orbits of nearby planetesimals and scattered them onto eccentric orbits. Gas drag acting on planetesimals with asteroid belt-crossing orbits decreased their eccentricities, causing many to become trapped on stable, lower-eccentricity orbits in the belt. Scattered objects originated from across the outer Solar System (out to 10-20 AU) and were preferentially trapped in the outer belt.  The belt's radial structure can be matched by associating implanted planetesimals with C-types and assuming that a small amount of planetesimals native to the belt represent the S-types.  The giant planets' growth also scatters objects onto terrestrial planet-crossing orbits, providing a potential source of water for Earth~\citep{raymond17}. The efficiency with which planetesimals are scattered toward the terrestrial region is higher when gas drag is weaker and thus increases in efficiency as the disk dissipates (as well as for larger planetesimals). This process is universal and happens any time a giant planet forms (meaning that it happened several times in the Solar System). The asteroid belt's excitation can be explained by processes such as chaotic excitation~\citep{izidoro16} or by secular excitation during the giant planet instability~\citep{deienno18}.  

An extreme version of the Low-mass Asteroid Belt model invokes a completely empty belt in which absolutely no planetesimals formed between Earth's and Jupiter's orbits~\citep{raymond17b}.  Under that assumption, the terrestrial planets' orbits are naturally reproduced, and enough planetesimals are scattered out from the terrestrial planet region and implanted in the main belt to account for the total mass in S-types. Given that the giant planets' growth invariably contributes C-types~\citep{raymond17}, the `Empty Asteroid belt' model thus proposes that all asteroids are refugees, implanted from across the Solar System.  However, there is a problem with the Empty Asteroid belt model in that S-type asteroids~\citep[associated with ordinary chondrites][]{bus02} are compositionally distinct from Earth~\citep[e.g.,][]{warren11}. In addition, the initial conditions for the simulations of \cite{raymond17b} essentially invoke a single generation of planetesimals to explain the terrestrial planets and S-types. However, measured ages imply that non-carbonaceous objects formed in several generations over the disk's lifetime~\citep[e.g.,][]{kruijer17}. This problem may in principle be solved if another generation of planetesimals formed past Earth's orbit but still interior to the asteroid belt. 

It is interesting to note that the implantation of planetesimals from the terrestrial planet forming region into the asteroid belt happens regardless of the formation model.  In the Empty Asteroid belt model this represents the main source of volatile-depleted asteroids~\citep{raymond17b}.  However, classical model simulations have also found that terrestrial planetesimals are implanted into the main belt, and with a similar efficiency~\citep{bottke06,mastrobuono17}.  This suggests that the present-day belt {\em must} contain a population of leftovers of terrestrial planet formation. It remains to be understood whether meteorites from such objects already exist in our collection or whether for unlucky reasons they are extremely rare (e.g., if there have not been any recent breakups of such asteroids).

The Low-mass Asteroid Belt model's weakest point is its initial conditions. Planetesimal formation models in the context of dust concentration and streaming instability in disks with realistic structures struggle to produce planetesimal distributions consistent with the Solar System. Some models do produce rings of planetesimals at $\sim 1$~AU suitable for terrestrial planet formation but no outer planetesimals that may have produced the giant planets' cores~\citep{drazkowska16,surville16}.  Other models produce planetesimals in outer planetary systems -- in particular just past the snow line -- but none in the terrestrial region~\citep{armitage16,carrera17,drazkowska17}. It remains unclear what conditions or processes are needed to produce planetesimal disks that are plausible precursors to the Solar System. The abundance of new studies shows that this issue may be resolved in the near-term. From a geochemical standpoint, if the building blocks of the terrestrial planets were concentrated in a narrow ring then it is difficult to understand observed differences between Earth and Mars and also why the Earth's chemistry is consistent with accretion from a heterogenous reservoir of material~\citep{rubie11}.

The Low-mass Asteroid Belt model is robust to a modest degree of orbital migration of Jupiter and Saturn. Once the terrestrial ring of planetesimals has formed it is mostly separated from the giant planets' dynamical influence.  In addition, there are disk-planet configurations for which Jupiter and Saturn's migration is slow or negligible~\citep{morby07a,pierens14}.

{\bf The Grand Tack model} invokes large-scale migration of Jupiter to sculpt the inner Solar System~\citep{walsh11}. The disk of rocky embryos and planetesimals is assumed to have extended smoothly from 0.5-0.7 AU out to the giant planet-forming region. Jupiter is assumed to have formed at 3-4 AU, opened a gap and started to type II migrate inward, shepherding and scattering the rocky bodies in its path~\citep[as in pervious studies focusing on exoplanets; ][]{fogg05,raymond06c,mandell07}. Meanwhile, Saturn grew on an exterior orbit and migrated inward in the very rapid, gap-clearing type III regime~\citep{masset03}. Saturn became locked in Jupiter's exterior 2:3 resonance in a shared gap in the disk. This configuration changes the balance of disk torques felt by the coupled Jupiter-Saturn system, causing both planets to migrate {\em outward} while maintaining the resonance~\citep{masset01,morby07a,pierens08,zhang10,pierens11,pierens14}. If Jupiter's turnaround, or ``tack'' point was at 1.5-2 AU then the disk of rocky material was truncated at $\sim1$~AU, creating an edge reminiscent of the outer edge of embryos in the Low-mass Asteroid Belt model.  Jupiter and Saturn's migration continued until the disk itself started to dissipate, stranding the planets on resonant orbits consistent with the later giant planet instability~\citep[see][]{nesvorny12}.

The disk of rocky material sculpted by Jupiter's migration can match the terrestrial planets' orbital and mass distributions and formation timescales~\citep{walsh11,jacobson14,jacobson14b,brasser16}. Despite having traversed the asteroid belt twice, the belt is re-populated by both scattered inner disk material (linked here with S-types) and implanted outer disk planetesimals (linked with C-types) and its orbital distribution and total mass match the present-day belt~\citep{walsh11,walsh12,deienno17}. During the giant planets' outward migration, some C-type material is scattered inward {\em past} the asteroid belt onto terrestrial planet-crossing orbits, providing Earth with the appropriate water budget with the correct isotopic signature~\citep{walsh11,obrien14} as well as atmophile elements~\citep[e.g., H, C, N, O and the noble gases;][]{matsumura16}.  The Grand Tack can also match the terrestrial planets' compositions thanks to the early mixing of material originally formed over several AU during Jupiter's inward migration phase~\citep{rubie15}.

The Grand Tack's weakness is the outward migration mechanism. Hydrodynamical simulations of Jupiter-mass and Saturn-mass planets embedded in isothermal disks universally find that the two planets become locked in 3:2 resonance and migrate outward~\citep{masset01,morby07a,pierens08,zhang10,pierens11}. Outward migration can extend across long stretches of planet-forming disks~\citep{crida09}. When the disk mass and viscosity are varied and a more realistic thermal structure is accounted for, there is a spectrum of evolutionary pathways for the gas giants' orbits~\citep{pierens14}. In disks with relatively low masses~\citep[smaller than the minimum-mass solar nebula model of][introduced in \S 1.2]{weidenschilling77,hayashi81} and moderate viscosities~\citep[viscous stress parameters of $10^{-4} \lesssim \alpha \lesssim 10^{-2}$, where $\alpha$ is the poorly-constrained parameter that controls the rate at which the disk evolves via angular momentum transport;][]{shakura73} the gas giants are trapped in 2:1 resonance and maintain roughly stationary orbits. In somewhat more massive disks with moderate viscosities, Jupiter and Saturn are trapped in 3:2 resonance and migrate outward as in the isothermal case. In very low-viscosity disks ($\alpha \approx 10^{-5}$) Jupiter and Saturn are trapped in 2:1 resonance but migrate outward~\citep{pierens14}. This diversity of potential outcomes encompasses the regimes appropriate for the three models presented in this section.

An additional issue is whether outward migration of Jupiter and Saturn can be maintained in the face of gas accretion~\citep[see discussion in][]{raymond14c}. In isothermal disks outward migration happens when two conditions are met~\citep{masset01}. First, Saturn must be at least half of its present-day mass in order to open a partial gap in the disk. Second, the Jupiter-to-Saturn mass ratio must be between 2 and 4. The outward migration of Jupiter and Saturn envisioned in the Grand Tack spans a wide range in orbital distance (from $\sim$1.5 to $>$5 AU for Jupiter) and takes a significant amount of time (likely on the order of 0.5-1 Myr for the entire outward migration phase). The question is, can the gas giants maintain the requisite conditions for outward migration during this entire phase when gas accretion onto the planets is taken into account? A definite answer to this question requires a good understanding of the disk's structure and evolution and of the planets' gas accretion.  Both of these processes remain too poorly understood to allow for a clear evaluation at this point, although given the large amount of interest in the Grand Tack there is hope for progress in the near-term.

{\bf The Early Instability model} proposes that terrestrial planet formation was strongly affected by the giant planet instability~\citep{clement18}. This constrains the instability to have taken place within $\sim 10$~Myr of the disk's dissipation~\citep{morby18} in order to have an impact on stunting Mars' growth~\citep{nimmo07,dauphas11}. Within the inner Solar System, an early instability effectively causes a rapid transition from a dynamically calm state to an excited one that bears a strong resemblance to the EEJS (`Extra Eccentric Jupiter and Saturn') configuration discussed in \S 4.2~\citep[from][]{raymond09c}. An early instability thus explains how the giant planets could have reached EEJS-like orbits in a self-consistent way at an early enough time to make a difference for terrestrial accretion.

Simulations by \cite{clement18,clement18b} show that an early instability acts to excite the asteroid belt and to clear out the Mars zone. This is in contrast to simulations of a late giant planet instability that often over-excite the terrestrial planets' orbits~\citep[or destabilize them entirely; ][]{brasser13,roig16,kaib16}. With an early instability, the depletion is so strong that a significant fraction ($\sim 20\%$) of simulations leave Mars' present-day orbital region completely empty and the median Mars analog is close to Mars' true mass. In many simulations planets grow much larger than a Mars mass at Mars' orbital distance, and are then excited to significant eccentricities by the instability, and then collide at pericenter with the growing Earth or Venus. Mars analogs in these simulations are often stranded embryos that avoid colliding with the larger embryos when these are scattered inward.  

The Early Instability model can match the asteroid belt and Earth's water content. The belt is pre-excited before the instability by scattering of planetesimals by resident embryos~\citep[][]{petit01,chambers01b,obrien07}. During the instability the belt is depleted~\citep[][]{morby10,roig15,clement18b} and embryos are removed, and surviving planetesimals provide a decent match to the present-day belt's eccentricity and inclination distributions~\citep{clement18,clement18b}.  As in the classical model, the belt's S-/C-type compositional dichotomy is assumed to be matched due to previous events~\citep[and recall that Jupiter's growth invariably implants C-types into the outer main belt][]{raymond17}. Water-rich material is delivered to Earth by the same mechanism as in the classical model~\citep{morby00,raymond07a,raymond09c}, via impacts from water-rich planetesimals and embryos originating in the outer main belt.  

An appealing aspect of the Early Instability model is that it simplifies the Solar System's timeline.  Rather than invoking separate mechanisms to sculpt the terrestrial- and giant planet systems, a single key event can explain them both. In addition, the terrestrial planets are best reproduced when the giant planets reach their actual configuration~\citep[as measured by the giant planets' orbits and the strength of different secular modes][]{clement18}.

The Early Instability model's weak points are related to the timing of the instability. Xenon isotopes are the first issue. Xenon in Earth's atmosphere is fractionated due to hydrodynamic escape and is enriched in heavier isotopes relative to Xenon found in chondritic meteorites or in the Solar wind~\citep{ozima02}. When this fractionation is taken into account, the isotopic signature of presumably primordial atmospheric Xenon is still significantly different from that of Xenon in the mantle. This suggests that another source of atmospheric Xenon had to exist with a different isotopic signature than chondrites or the Solar wind. Comet 67P/Churyumov-Gerasimenko matches the missing Xenon signature and Earth's atmosphere can be matched with a 20-80 mixture of cometary and chondritic Xenon~\citep{marty17}.  In contrast, the mantle's Xenon was purely chondritic~\citep[e.g.,][]{sujoy12,caracausi16}. The long-standing mystery of Earth's Xenon thus appears to be solved if Earth's atmospheric Xenon included an additional cometary component. In this scenario the mantle's Xenon came from the inner Solar System as Earth was accreting. The atmosphere's Xenon came from a late bombardment of comets necessarily related to the giant planet instability, which represent a significant source of noble gases but not of water~\citep{marty16}. This elegant argument relates the relative timing of Earth's accretion and the cometary bombardment linked with the giant planet instability.  In principle, a very early instability would cause a cometary bombardment at the same time as accretion such that one might expect Earth's mantle Xenon to have the same isotopic signature as its fractionation-corrected atmospheric Xenon.  In contrast, a late cometary bombardment would only affect the atmospheric Xenon signature.  

At face value, the Xenon constraint would seem to rule out a giant planet instability earlier than the Moon-forming impact. The relative timing of the instability and Moon-forming impact affects the ability of all terrestrial planet formation models to match the Xenon constraint, not just the Early Instability model. However, there are two caveats. First, the particular Xenon signature has only been measured in a single comet (although it is the only comet in which the signature could have been detected). Second, the repartition of the Xenon signature between the mantle and atmosphere during Earth's impact-driven evolution is uncertain and certainly depends on factors related to Earth's growth history and chemical evolution.

Another potential conflict comes from the orbital structure of the Kuiper belt. \cite{nesvorny15} found that the high-inclination classical Kuiper belt can be matched by a smooth phase of migration of Neptune that was interrupted by the instability. If true, this would restrict the earliest possible timing of the instability to be $\sim$~20 Myr after gas disk dissipation, too late to stunt Mars' growth. This constraint is relatively indirect, as other models may potentially explain the high-inclination population. 

\bigskip
\noindent
\textbf{4.5 An outstanding issue: the mass deficit in the very inner Solar System}
\bigskip

The very low amount of mass in the very inner Solar System (interior to Mercury's orbit) is hard to understand. This feature is important because it represents a divide between the Solar System and exoplanet systems. Roughly half of all stars have Earth-sized or larger planets interior to Mercury's orbit~\citep[e.g.][]{howard10,howard12,mayor11}. However, our terrestrial planets are consistent with having formed from a narrow ring of planetesimals and embryos between Venus' and Earth's present-day orbits~\citep{hansen09,walsh16,raymond17b}. Models such as those presented in \S 4.4 can explain the outer edge of this ring of rocky material, for example by invoking dynamical truncation by the migrating Jupiter~\citep{walsh11}. However, the inner edge of this ring -- and the absence of other planets closer-in than Mercury -- remains challenging to explain. Of course, given its very large iron core and the reduced oxidation state of its crust and mantle, Mercury itself is a challenge to explain~\citep[e.g.,][]{ebel17}. 

It was proposed by \cite{leake87} that a population of planetesimals formed on orbits interior to Mercury's and later bombarded Mercury. Indeed, there is a belt of dynamically stable orbits between 0.06 and 0.21 AU~\citep[sometimes called the ``Vulcanoid zone''; ][]{evans99}. However, a belt of planetesimals on such close-in orbits would undergo vigorous collisional grinding~\citep{stern00}. Efficient removal of small bodies via radiative transport would remove the bulk of the population's mass, and the surviving planetesimals would themselves be further depleted by Yarkovsky effect-driven drift into unstable orbital configurations~\citep{vokr00}. Very few km-scale planetesimals are expected to survive. Only planetesimals large enough to have significant self-gravity ($D \gtrsim 100$~km) would have survived, and to date none have been found. This suggests that a belt of 100 km-scale planetesimals typical of the streaming instability~\citep{simon16,schafer17} did not form closer-in than Mercury or, if it did, it was dynamically -- not collisionally -- removed.

\cite{ida08} proposed that the rocky mass interior to roughly Venus' orbit was removed by inward migration. They assumed that accretion proceeds roughly as a wave sweeping outward in time and that the large embryos produced by accretion migrated inward and fell onto the young Sun. Embryos massive enough to undergo gas-driven orbital migration only had time to form interior to roughly Venus' orbit. While appealing, this model ignores the fact that disks have inner edges~\citep[as demonstrated by simulations of magnetic accretion onto young stars; see][]{romanova04}. Embryos can migrate to the inner edge of the disk, where they are trapped by a strong positive torque~\citep{masset06,romanova06}, but they remain a great distance from the surface of the star (roughly an order of magnitude larger than the stellar radius).

\cite{raymond16} proposed that {\em outward} migration could explain the very inner Solar System's mass deficit. Assuming planetesimals to form throughout the inner disk, they invoked the rapid formation of a large core of a few Earth masses close to the Sun. This object could plausibly have formed by trapping a fraction of inward-drifting pebbles, perhaps at a pressure bump in the very inner disk such as found in the disk models of \cite{flock17}. For objects of a few Earth masses embedded in the inner parts of radiative viscous disks, migration is often directed outward~\citep[the details depend on parameters such as the disk metallicity and accretion rate; see ][]{bitsch15}. The large core's outward migration through a population of planetesimals and embryos is analogous to the case of a large planet migrating inward through similar objects discussed in \S 3.2~\citep[see also][]{fogg05,raymond06c,izidoro14}. For outward migration timescales of $\sim 10^5$ years, \cite{raymond16} found that the core shepherds embryos and planetesimals in exterior resonances and clear out the inner Solar System. However, this mechanism typically broke down when the core reached 0.5-1 AU due to scattering between shepherded bodies. The core's migration would have continued to a zero-torque location past the snow line~\citep{bitsch15}, often contaminating the primordial asteroid belt with material from the very inner Solar System. In the context of this model, the migrating core represents Jupiter's core and therefore it is expected to contain a large fraction of rock. The weakness of this idea is the setup, as it requires 1) a large, close-in core that forms much faster than more distant, smaller embryos, and 2) a disk in which migration is directed {\em outward}. Yet this setup (at least point 1) appears to be plausible as many disk models have an inner pebble/dust trap~\citep[e.g.][]{flock17} and this idea is at the heart of the drift model for super-Earth formation~\citep{chatterjee14}. 

\cite{morby16} proposed that the edge in the presumed disk of rocky building blocks at 0.7 AU corresponds to the location of the silicate condensation line at early times when the disk was hot.  Rocky planetesimals that formed early could not have formed closer than the silicate condensation line. That early generation of planetesimals continued to grow by mutual collisions or by accreting pebbles but the lack of closer-in material would be preserved if no other planetesimals formed closer-in. There are two main uncertainties in this model.  First, why did no planetesimals form closer-in at later times? Given that several generations of planetesimals are thought to have formed in the inner Solar System~\citep[e.g.,][]{kruijer17}, it remains to be understood why none would form closer-in.  Second, what happens to the pebbles that drift inward past the growing planetesimals?  Tens of Earth masses in pebbles are likely to have drifted inward past the rocky planetesimals~\citep[e.g.,][]{lambrechts14}. These are generally assumed to have reached the inner parts of the disk and simply sublimed. If even a small fraction is trapped, then it could lead to the formation of a large core as in the \cite{raymond16} model.

Two papers proposed that the early Solar System contained a population of super-Earths that were destroyed. \cite{volk15} proposed that very energetic collisions ground the planets to dust. However, examination of the collisional parameters in the simulated collisions suggest that they are far below the catastrophic destruction threshold~\citep{leinhardt12,wallace17} and it is hard to understand how all of the planets' mass could have been removed. Unlike km-scale planetesimals -- which, as discussed above, would indeed be ground to dust and removed on orbits closer-in than Mercury's~\citep{stern00,vokr00} -- planets' self-gravity prevents their total destruction.  

In contrast, \cite{batygin15} invoked collisional debris generated by the Grand Tack to push a population of primordial super-Earths onto the young Sun. While there are key issues related to the mechanism at play~\citep[see discussion in][]{raymond16}, the main problem with this model is that, as described above, planets cannot simply migrate onto their stars. Rather, their migration is blocked by the inner edge of the disk~\citep{masset06,romanova06} and their evolution should be similar to the {\it breaking the chains} model~\citep[see also Fig.~\ref{fig:exo} and discussion in \S 3.1]{izidoro17,izidoro18b}.

To conclude, in our minds the origin of the mass deficit closer-in than Venus' orbit remains unexplained. While successful models to match the deficit do exist, there is no clear theory that does not have significant counterarguments or require specific assumptions. Based on our current understanding it is not plausible to invoke migration onto the Sun as a mechanism for losing close-in material.  Rather, close-in material may have been swept outward by a migrating core~\citep{raymond16}, or the inner disk may simply never have produced planetesimals~\citep{morby16}.  It is also entirely possible that another mechanism may be responsible.

\section{\textbf{EXTRAPOLATION TO EXO-EARTHS: FORMATION TIMESCALES AND WATER CONTENTS}}

We now turn our attention to the more general question of Earth-like planets around other stars.  Just how `Earth-like' should we expect Earth-sized planets in the habitable zones of their host stars to be?  And how does this depend on other properties of these systems (in particular observable ones) such as the planetary system architecture, the planet size/mass and the stellar type? In this section we address the formation and water contents of potentially habitable planets. We do not address the question of what conditions are needed for habitability. Rather we simply assume that $\sim$~Earth-mass planets in the habitable zones of their parent stars are viable candidates.


The Solar System's terrestrial planets are thought to have accreted in steps (see \S 2). First, planetesimals formed from drifting pebbles and dust. Then planetesimals grew into planetary embryos by accreting pebbles and other planetesimals. Embryos grew slowly enough and remained small enough not to have undergone any significant migration. Finally, there was an extended phase of giant collisions between embryos lasting $\sim 100$~Myr.

To generalize the formation of Earth-mass planets, we want to know how universal each of these steps is. Do all planet-forming disks follow the same general pattern as ours?  

While great strides have been made in understanding how 100 km-scale planetesimals form, models disagree on where and when they form~\citep[e.g.,][]{drazkowska16,carrera17,drazkowska17}. We can imagine that planetesimal disks might be quite diverse in their structures; for instance, the Low-mass Asteroid Belt model is based on a particular structure (see \S 4.3). However, given our limited understanding of planetesimal formation in the Solar System we cannot reasonably extrapolate to other systems.  For the purpose of this discussion we will simply assume that planetesimal formation is robust and has no strong radial dependence.

Embryo growth is a critical step. The Solar System's terrestrial planets are consistent with having formed from a population of $\sim$~Mars-mass embryos~\citep[e.g.][]{morby12b}. The largest planetesimals undergo runaway accretion (of other planetesimals) and become Moon- to Mars-mass embryos~\citep[e.g.][]{greenberg78,wetherill93,kokubo00}. For a minimum-mass disk embryos take 0.1-1 Myr to grow at 1 AU, at which point they excite the orbits of nearby planetesimals, decrease the effects of gravitational focusing and their growth from planetesimal accretion slows down drastically~\citep[e.g.][]{kokubo98,leinhardt05}. Yet pebble accretion should continue and even accelerate~\citep[e.g.][]{ormel10,lambrechts12}. 

Matching the terrestrial planets therefore requires a quenching of pebble accretion to prevent embryos at 1 AU from growing too massive~\citep[see][for simulations of the pebble flux-governed bifurcation between terrestrial planets and rocky super-Earths]{lambrechts18}. This may have happened as a consequence of the growth of Jupiter's core to the {\em pebble isolation mass}, at which point it created a barrier to inward pebble drift~\citep{morby12,lambrechts14b,bitsch18}. \cite{kruijer17} used the temporal co-existence of meteorites with different nucleosynthetic signatures (carbonaceous vs. non-carbonaceous) to infer that Jupiter's core did indeed provide a barrier within 1 Myr after CAIs~\citep[see also][]{desch18}. From that point on, pebble accretion was shut off in the inner Solar System and the terrestrial planets grew by accreting planetesimals and embryos.

This line of thinking implies that the timing of the growth of Jupiter's core was critical~\citep[see, e.g.][]{bitsch15b,bitsch18b}. If it had grown much more slowly, pebble accretion would have generated more massive terrestrial embryos. These large embryos would then have migrated and likely followed the {\em breaking the chains} evolution discussed in \S 3.1 (see also Fig.~\ref{fig:exo}). It is not clear that faster growth of Jupiter's core would have had much of an effect, as pebble accretion for sub-Mars-mass embryos is relatively slow. If we assume that the growth of large outer cores varies significantly from disk to disk, these differences in timing can have big consequences. Systems with fast-growing cores may preserve their small inner rocky embryos, whereas in systems with slower-growing cores terrestrial embryos grow sufficiently fast that they cannot avoid migration. 

Migration must play an important role in the formation of many habitable zone planets. This happens if a) an outer core (analogous to Jupiter's core) formed slowly enough for large embryos to grow by pebble accretion, or b) the central star is low enough in mass that the formation timescale in the habitable zone is very short. The accretion timescale depends on the local disk surface density (in planetesimals) and the orbital timescale~\citep{safronov69}. Given the strong scaling of the habitable zone with stellar type~\citep[because of the strong stellar mass-luminosity scaling; ][]{scalo07,mulders15c} the accretion timescales for planets in the habitable zones of low-mass stars are much shorter than for Sun-like stars~\citep{raymond07b,lissauer07,dawson15}. Extrapolating from Earth's 50-100 Myr formation timescale, planets in the habitable zones of stars less massive than 1/2 to 1/3 of a Solar mass should form quickly, with most of the assembly taking place during the gas disk phase even in the absence of pebble accretion~\citep{raymond07b}. 

In these systems embryos should follow the {\em breaking the chains} behavior described above: they should migrate into resonant chains anchored at the inner edge of the disk, most of which go unstable when the gas dissipates. Habitable zone planets would undergo a final phase of giant impacts shortly after disk dispersal~\citep{ogihara09,izidoro17,izidoro18b}. The compositions of planets in these systems would vary~\citep{raymond18b}. Some could be made up entirely of large rocky embryos. However, given that embryos should still eventually grow large past the snow line, some planets would likely contain a significant fraction of their mass in volatiles. 

What about systems with more massive central stars in which an outer core forms quickly? Fast-growing cores are themselves likely to migrate. If the disk properties are such that there is a zero-torque migration trap that lasts for close to the entire disk lifetime~\citep{lyra10,bitsch15}, the inner planetary system may be protected from the core's migration. In some cases cores may accrete gas to become gas giants and transition to slower, type 2 migration. In those cases terrestrial planet formation should in principle follow the same pattern as in the Solar System provided the giant planets do not migrate all the way into the habitable zone~\citep[see][]{fogg05,raymond06c}. When migration is not stopped, the large core would plow into (or through, depending on the migration timescale) the growing terrestrial planets~\citep{izidoro14,raymond18b}. Planets can still form in the habitable zones in such systems but they are not rocky worlds like Earth. Rather, such planets should have compositions representative of the region past the snow line, presumably with large water contents. 

Some simple analysis can hint at the distribution of outcomes. Low-mass (M) stars are the most common by number~\citep[e.g.,][]{chabrier03}. The disk mass is observed to be a steeper than linear function of the stellar mass~\citep[scaling roughly as $M_{disk} \propto M_\star^{1.6}$ albeit with large scatter; e.g.,][]{scholz06,pascucci16}. For low-mass stars, the snow line is farther-removed from the habitable zone than for FGK stars, as measured simply by the snow line to habitable zone distance ratio~\citep{mulders15c}. Given that the growth timescale scales with the disk mass~\citep[e.g.,][]{safronov69} we expect outer cores to grow slowly around low-mass stars. Assuming that there is sufficient local material to build an Earth-mass planet in situ in the habitable zone, then rocky embryos should form quickly~\citep{raymond07b,lissauer07b,dawson15},  undergo pebble accretion, and migrate. Embryos from past the snow line are also likely to migrate later in the disk lifetime, as disks around low-mass stars are observed to have longer lifetimes than around Sun-like stars~\citep{pascucci09}. The location of the disk's inner edge depends on the rotation rate of young stars and is unlikely to be a strong function of the stellar mass. This means that the habitable zones of low-mass stars are closer to the inner edge of the disk than those of higher-mass stars. Rocky planets that form near the habitable zone do not necessarily migrate far away. Given the late instabilities characteristic of {\em breaking the chains} behavior, the final planets should be a mixture of embryos that started with terrestrial compositions and those with ice-rich compositions from past the snow line.  This should result in a diversity of planetary compositions, from pure rock planets to planets with tens of percent ice by mass~\citep[neglecting various water/ice loss processes; e.g., ][]{grimm93,genda05,marcus10,monteux18}. 

The main difference between this scenario for low-mass and higher-mass stars is the accretion timescale in the habitable zone~\citep{raymond07b}. For FGK stars, terrestrial embryos in the habitable zone are less likely to grow fast enough to migrate if they only accrete planetesimals. Compared with the same setting around low-mass stars, growing rocky planets in the habitable zones of FGK stars are more likely to be protected from pebble accretion by a fast-accreting core, and to have a large ice-rich core migrate into the terrestrial zone. Given the much faster core accretion compared with terrestrial accretion, close-in planets that result from the inward migration of large cores are likely to have higher average volatile contents than for low-mass stars. Of course, in some situations outer cores will accrete into gas giants and (in some cases) remain on wide orbits. In these cases the terrestrial planets' accretion is protected, although the gas giants' growth may shower the terrestrial zone with volatile-rich planetesimals~\citep{raymond17}. 

Compared with FGK stars, low-mass stars are found to have more super-Earths smaller than $2 \rearth$ but fewer sub-Neptunes between 2 and $4 \rearth$ and a higher total average planet mass on close-in orbits~\citep{mulders15,mulders15b}. This can be explained by the reasoning presented above if the smaller super-Earths preferentially formed from migrating rocky embryos and larger sub-Neptunes formed mainly from migrating ice-rich embryos. Higher-mass stars have more gas giants~\citep{johnson07,lovis07}. While growing gas giant cores block pebbles from drifting past~\citep{morby12,lambrechts14b,bitsch18}, gas giants themselves block embryos from migrating past~\citep{izidoro15a}. This might tilt the scales in favor of low-mass stars having a higher average total mass in close-in planets.

Let's put the pieces together. Given that M stars dominate by number, the formation pathway of their habitable zone planets likewise dominates. The habitable zones of low-mass stars are so close-in that we expect planets to grow rapidly from both rocky and ice-rich material and to follow the {\em breaking the chains} evolution described in \S 3.1. The bulk of these planets' growth took place during the gas disk phase, with last giant impacts happening during a late instability shortly after the dissipation of the disk. Their compositions are likely to span a wide range from pure rock to ice-rich depending on the objects' individual growth histories. 

Earth-mass planets in the habitable zones of FGK stars are likely to have followed one of two pathways. In systems in which cores of $10-20 \mearth$ or more grow quickly past the snow line, the flux of pebbles toward the inner system is shut off. In some systems (like our own) these large cores grow into gas giants, which migrate slowly and may remain isolated from the terrestrial zone. Further growth involves planetesimal and embryo accretion, and rocky embryos are unlikely to reach high enough masses to migrate within the gas disk's lifetime. In other systems outer, ice-rich cores do not become gas giants but instead migrate inward into the terrestrial zone. In that case habitable zone planets may typically be very volatile-rich~\citep[e.g.][]{kuchner03}. 

How can we use observations to constrain these ideas? Systems that follow the {\em breaking the chains} evolution should commonly form habitable zone planets. However, the habitable zone planets themselves should have a diversity of compositions, encompassing systems in which rocky embryos grew large enough to migrate and those in which they did not. Low-mass stars should be more likely to have migrating rocky embryos and, since their habitable zones are likely to be closer to the inner edge of the disk, are also more likely to retain rocky embryos/planets in the habitable zone.  Finally, systems with outer giant planets~\citep[beyond 2.5-3 AU for Sun-like stars; see][]{raymond06d} and no inner giants are the best candidates for having habitable zone planets with small but non-zero, Earth-like water contents.

\section{\textbf{DISCUSSION}}

\bigskip
\noindent
\textbf{6.1 Central processes that sculpt planetary systems}
\bigskip

Two processes appear to be ubiquitous in planet formation: migration and instability. These are essential ingredients in explaining the origins of exoplanet systems as well as the Solar System. The {\em breaking the chains} model proposed by~\cite{izidoro17,izidoro18b} invokes inward migration of large embryos into long chains of mean motion resonances anchored at the inner edge of the disk. The vast majority of resonant chains become unstable when the disk dissipates, leading to a late phase of giant collisions. Given that virtually all observed super-Earths are massive enough to undergo gas-driven migration, we argued in \S 3.1 that, regardless of when and how they form, all super-Earth systems converge to the {\em breaking the chains} evolution.  

The population of giant exoplanets may be explained by invoking the formation of multiple gas giants that also migrate into compact resonant configurations (see \S 3.2). As for super-Earths, the vast majority of systems undergo instabilities when the disk dissipates or perhaps even during the late phases of the disk lifetime. The outcome of an instability correlates with the Safronov number, which is the ratio of the planets' escape speed to the local escape speed from the system~\citep{safronov69,ford08}. For high Safronov numbers the planets impart strong enough gravitational kicks that scattering is favored over collisions. Giant planet instabilities thus lead to planet-planet scattering and the surviving planets match the observed giant exoplanet eccentricity distribution~\citep[as well as the correlated mass-eccentricity distribution; ][]{raymond10}.

Solar System formation models likewise invoke different combinations of migration and instability.  All current evolutionary pathways rely on the {\em Nice model} instability in the giant planets' orbits, although the timing of the instability is uncertain~\citep{morby18}. The Grand Tack model uses a Jupiter's specific migration path to deplete the asteroid belt and Mars' feeding zone~\citep{walsh11}. In contrast, the Early Instability model proposes that an early giant planet instability was responsible for depleting the asteroid belt and Mars region~\citep{clement18}. 

Two additional processes are central in setting the stage: planetesimal formation and pebble accretion. They determine where and when embryos large enough to migrate can form. The Low-Mass Asteroid belt model proposes that the mass depletion in the Mars- and asteroid belt regions was inherited from the planetesimal formation stage~\citep{drazkowska16}. Likewise, the timing of the formation of Jupiter's core -- which acted to block the flux of pebbles to the inner Solar System -- was a key moment in keeping the terrestrial planets {\it terrestrial} (see discussion in \S 5). 

Different processes have different philosophical implications. Migration and instability both act to erase the initial conditions. Many formation pathways lead to a phase of migration, but once a system migrates it starts to forget its initial conditions, as all pathways converge to the same evolution. It is for that reason that density constraints on super-Earth compositions are so important: they are our only clue as to the planets' origins, and even for accurate measurements density is a weaker diagnostic than one would like (see discussion in \S 3.1). Likewise, the chaotic nature of instabilities makes it impossible to rewind the clock on a system of planets with eccentric orbits to uncover their pre-instability configuration~\citep[although statistical studies try to do just this; e.g., ][]{nesvorny12}.

In contrast, bottom-up accretion retains a memory of its initial conditions. The planets that form from a disk with a given surface density still follow that same profile~\citep{raymond05}. This is the central argument behind the `minimum-mass solar nebula' model~\citep{weidenschilling77,hayashi81}. This allows us to put strong -- but not unique -- constraints on the properties of the precursor disks that formed systems of small planets, assuming that accretion was the main process involved.  For instance, a few different initial distributions of planetesimals and embryos can match the terrestrial planets but they all share common properties such as a strong mass deficit past Earth's orbit and interior to Venus'~\citep{hansen09,izidoro15c}.  

How well do planetesimal formation and pebble accretion remember their initial conditions?  In the current paradigm, planetesimals form when drifting dust and pebbles are sufficiently concentrated to trigger the streaming instability~\citep{johansen14}. Exactly where and when this happens depends on the underlying disk model~\citep[e.g.,][]{drazkowska16,carrera17}. However, once planetesimal formation is triggered, objects form with a characteristic size distribution~\citep{simon17,schafer17}. Thus, while accretion may preserve a trace of where planetesimals formed and in what abundance, planetesimal formation itself does not retain a memory of dust coagulation and drift. Pebble accretion is too parameter-dependent to retain a memory of the historical pebble flux. If the pebble size (or Stokes number) and the initial planetesimal/embryo mass were known this might be possible.

\bigskip
\noindent
\textbf{6.2 Planet formation pathways and bifurcations: How did our Solar System get so weird?}
\bigskip

There are a few key bifurcation points in planetary system formation. At these points, small differences in outcome lead to very different evolution. We consider the key bifurcation points to be: 1) disk properties, 2) planetesimal formation (where? when?), 3) giant planet formation, 4) instability trigger (timing). We now go over each of these bifurcation points, then discuss which pathway the Solar System must have followed.

A star's protoplanetary disk is its cradle, where its planetary system is born and raised.  The characteristics of the disk and its evolution are perhaps the single most important factor in planet formation. While the detailed structure and evolution of disks are themselves poorly understood~\citep[see][]{morbyraymond16}, observations and theory demonstrate that there is a diversity in disk mass, structure and lifetime~\citep[e.g.,][]{haisch01,williams11,bate18}. The disk mass may itself be the key determinant of planetary system evolution~\citep{greaves06,thommes08}. More massive disks should more readily form planetesimals, embryos and gas giants. Given their higher abundance of solids, higher-metallicity stars should also more readily form planetesimals and planets.

Where and when planetesimals form is of vital importance. If planetesimals form early then they are bathed in a flux of pebbles and can quickly grow into large embryos/cores and perhaps even gas giants~\citep[e.g.][]{lambrechts14,bitsch15b}. However, if planetesimals only form late -- perhaps triggered by the dissipation of the gas disk and the accompanying increase in dust/gas ratio~\citep{throop05,carrera17} -- then the bulk of their growth must appeal to gas-free processes such as planetesimal accretion and no gas giants can form.  The radial distribution of planetesimals is naturally of vital importance to planet formation. The Low-mass Asteroid Belt model relies on planetesimals forming in a narrow ring in the inner Solar System~\citep{drazkowska16} whereas the Grand Tack~\citep{walsh11} and Early Instability~\citep{clement18} models were devised assuming that planetesimals did indeed form in the Mars region and asteroid belt and that a depletion mechanism was needed.

The formation of a giant planet is essentially two bifurcation points. Once a planet's core reaches the pebble isolation mass~\citep[of roughly $20 \mearth$ at Jupiter's orbit for characteristic disk models;][]{lambrechts14b,bitsch18}, the pebble flux is blocked and the entire planetary system interior to the core is cut off from further pebble accretion. We argued in \S 5 that the timing of the formation of a pebble-blocking core relative to the growth of inner embryos is the central parameter that determines whether most super-Earths are likely to be rocky or ice-dominated. After a core undergoes rapid gas accretion to become a gas giant and carves a gap in the disk, it also blocks the inward migration of any other large cores that form on exterior orbits~\citep{izidoro15a}. This is a second way in which a wide-orbit planet cuts off its inner planetary system from the inward-drifting/migrating mass. Of course, if the wide-orbit planet migrates inward then it itself becomes that inward-migrating mass.

Triggering instability is the final and perhaps most dramatic bifurcation point. While instability appears to be near-ubiquitous, the impact of instabilities can vary. For instance, the Solar System's giant planets are thought to have undergone an instability but only a very weak one when compared with the instabilities in most exoplanet systems. Indeed, the instabilities characteristic of giant exoplanet systems often drive the growing terrestrial planets into their host star~\citep{raymond11,raymond12}. The late stage accretion of the terrestrial planets also represents a form of instability that concluded with the Moon-forming impact. The {\em breaking the chains}~\citep{izidoro17,izidoro18b} evolution characteristic of super-Earth systems causes much more dramatic late instabilities that involve collisions between much larger (typically $\sim 5 \mearth$) objects and, given their close-in orbits, much higher collision speeds. Of course, a small minority of systems avoid instability.  Stable systems easily recognized by their resonant orbits, which are systematically destroyed by instabilities~\citep[although scattering does generate resonances in a small fraction of cases; ][]{raymond08b} .

What path must the Solar System have taken with regards to these bifurcations?  The Sun's planet-forming disk may have been somewhat more massive than average, with enough mass to form the cores of several gas giants (which total $\sim 40-50 \mearth$), but not enough to form multiple Jupiter-mass planets.  A few Earth-masses worth of rocky planetesimals must have formed early in the terrestrial planet region, either in a smooth disk or in one or more rings. Planetesimals also formed beyond the snow line and produced the giant planets' cores. Jupiter's core grew fast enough to starve the inner Solar System of pebbles within $\sim 1 Myr$, keeping the precursors of carbonaceous and non-carbonaceous meteorites physically separated~\citep{kruijer17}, preventing further growth of the terrestrial planets' constituent embryos, and fossilizing the snow line~\citep{morby16}. Jupiter's growth also stopped the ice giants and Saturn's core from migrating into the inner Solar System~\citep{izidoro15a,izidoro15b}. When the disk dissipated, the inner disk of terrestrial embryos entered its late instability, which lasted $\sim 100$~Myr but during which Mars remained mostly isolated and protected. The giant planets' orbits became unstable sometime within the 500 Myr following disk dissipation. While the disk's dispersal is the main natural trigger for instability, some geochemical arguments (e.g., the atmospheric Xenon constraint; see \S 4.3) point to a later trigger.  While the instability did clear out the primordial Kuiper belt, it was far less dramatic than instabilities characteristic of extra-solar systems.

The Solar System's presumed evolution contains multiple unusual occurrences.  First, the gas giants' masses are quite different. The fact that the most massive giant exoplanets have the highest eccentricities~\citep{wright09} indicates that massive gas giants (of roughy $1 M_{Jup}$ or above) typically form in systems with other, roughly equal-mass gas giants that go unstable~\citep{raymond10}. Second, Jupiter's orbit remained wide of the terrestrial region. This may be because of the dynamical influence of Saturn; the Jupiter-Saturn system can migrate outward or remain on near-stationary orbits depending on the disk properties~\citep{pierens14}. However, avoiding inward migration depends on a Jupiter/Saturn mass ratio close to its current value~\citep{masset01}, so this unusual occurrence may be intrinsically linked with the previous one. Third, the giant planet instability did not include any close encounters between Jupiter and Saturn. Simulations show that such an encounter would likely have ejected Saturn and stranded Jupiter with an eccentricity of $\sim 0.2$~\citep{morby07}. In other words, Jupiter's eccentricity would be typical of giant exoplanets if its instability had proceeded in typical fashion. However, in that case the terrestrial planets may well have been driven into the Sun~\citep{raymond11}. 

We interpret these unusual occurrences as why our Solar System is {\em weird}. These evolutionary steps explain why Sun-Jupiter systems are rare within the known exoplanet sample (see \S 1.1). This in turn implies that terrestrial planet systems like ours are also rare, although an understanding of the timescales of different processes is needed to assess that assertion in a more careful way (see \S 5). 

\bigskip
\noindent
\textbf{6.3 A digression on the significance of models}
\bigskip

A successful formation model is expected to match a planetary system (or a distribution of systems) in broad strokes and with a suitable success rate. But what exactly constitutes the `breadth' of the `strokes' needed for success?  And at what rate is a model deemed successful? These inherently philosophical questions are central to models of planet formation. If a model matches the Solar System in 1\% of simulations, should we consider the problem solved? Or should we continue to test other models?  And if model A provides a match in 30\% of cases and model B in 10\%, can we be confident that model A is truly preferred over model B? 

We do not pretend to have a concrete solution, but we think it important to keep such considerations in mind.  We expect that in the future, global planet formation modeling may make use of more rigorous statistical methods to address these issues.

\bigskip
\noindent
\textbf{6.4 Paths Forward}
\bigskip

There remain a plethora of outstanding problems in planet formation.  As described in \S 4.4, studies will strongly constrain (and may falsify some) models of Solar System formation in the coming years. Nonetheless, we encourage the development of new models. NASA's OSIRIS-REX and JAXA's Hayabusa2 missions will return samples of carbonaceous asteroids (the B-type Bennu and the Cg-type Ryugu) that will certainly improve our understanding of the formation conditions of such objects and provide additional constraints on their origins. NASA's Lucy mission will study Jupiter's co-orbital asteroids, thought to have been captured during the giant planet instability~\citep{morby05}, and NASA's Psyche mission will study an apparently metallic asteroid that may have originated in the inner parts of a differentiated planetary embryo.  Meanwhile, upcoming exoplanet-focused instruments such as NASA's TESS and ESA's PLATO and ARIEL missions will deepen our understanding of the orbital architecture of planetary systems as well as their more detailed characteristics. This will provide additional constraints on exoplanet formation models. 

Among our ever-increasing stockpile of extremely valuable data, we conclude this chapter by emphasizing the need for global models to connect the dots. Models should not be constrained by dogma or current paradigms. Of course, a model is only viable if matches observations or measurements. A model is most useful if it is testable in the near term. And a model is most relevant when it lays the broadest possible foundation.  This means putting Solar System formation in the context of extra-solar planets.


\vskip .5in
\noindent \textbf{Acknowledgments.} \\
We thank referees John Chambers and Kevin Walsh for constructive reports, and are grateful to all of our colleagues who helped develop the ideas presented here. We each thank the Agence Nationale pour la Recherche for funding and support via grant ANR-13-BS05-0003-002 (grant MOJO). A.~I. acknowledges financial support from FAPESP (grants 16/12686-2 and 16/19556-7). S.~N.~R. also acknowledges NASA Astrobiology Institute's Virtual Planetary Laboratory Lead Team, funded under solicitation NNH12ZDA002C and cooperative agreement no. NNA13AA93A.


\end{document}